\documentclass[twoside,twocolumn,9pt]{article}
\usepackage{extsizes}
\usepackage[super,sort&compress,comma]{natbib} 
\usepackage[version=3]{mhchem}
\usepackage[left=1.5cm, right=1.5cm, top=1.785cm, bottom=2.0cm]{geometry}
\usepackage{balance}
\usepackage{widetext}
\usepackage{times,mathptmx}

\usepackage{sectsty}
\usepackage{graphicx} 
\usepackage{lastpage}
\usepackage[format=plain,justification=raggedright,singlelinecheck=false,font={stretch=1.125,small,sf},labelfont=bf,labelsep=space]{caption}
\usepackage{float}
\usepackage{fancyhdr}
\usepackage{fnpos}
\usepackage[english]{babel}
\usepackage{array}
\usepackage{droidsans}
\usepackage{charter}
\usepackage[T1]{fontenc}
\usepackage[usenames,dvipsnames]{xcolor}
\usepackage{setspace}
\usepackage[compact]{titlesec}
\usepackage{epstopdf}%This line makes .eps figures into .pdf - please comment out if not required.

\definecolor{cream}{RGB}{222,217,201}

\usepackage{amsmath}
\usepackage{amssymb}
\usepackage{hyperref}

\begin{document}

%\pagestyle{fancy}
%\thispagestyle{plain}
%\fancypagestyle{plain}{

%%%HEADER%%%
%\fancyhead[C]{\includegraphics[width=18.5cm]{head_foot/header_bar}}
%\fancyhead[L]{\hspace{0cm}\vspace{1.5cm}\includegraphics[height=30pt]{head_foot/journal_name}}
%\fancyhead[R]{\hspace{0cm}\vspace{1.7cm}\includegraphics[height=55pt]{head_foot/RSC_LOGO_CMYK}}
%\renewcommand{\headrulewidth}{0pt}
%}
%%%END OF HEADER%%%

%%%PAGE SETUP - Please do not change any commands within this section%%%
\makeFNbottom
\makeatletter
\renewcommand\LARGE{\@setfontsize\LARGE{15pt}{17}}
\renewcommand\Large{\@setfontsize\Large{12pt}{14}}
\renewcommand\large{\@setfontsize\large{10pt}{12}}
\renewcommand\footnotesize{\@setfontsize\footnotesize{7pt}{10}}
\makeatother

\renewcommand{\thefootnote}{\fnsymbol{footnote}}
\renewcommand\footnoterule{\vspace*{1pt}% 
\color{cream}\hrule width 3.5in height 0.4pt \color{black}\vspace*{5pt}} 
\setcounter{secnumdepth}{5}

\makeatletter 
\renewcommand\@biblabel[1]{#1}            
\renewcommand\@makefntext[1]% 
{\noindent\makebox[0pt][r]{\@thefnmark\,}#1}
\makeatother 
\renewcommand{\figurename}{\small{Fig.}~}
\sectionfont{\sffamily\Large}
\subsectionfont{\normalsize}
\subsubsectionfont{\bf}
\setstretch{1.125} %In particular, please do not alter this line.
\setlength{\skip\footins}{0.8cm}
\setlength{\footnotesep}{0.25cm}
\setlength{\jot}{10pt}
\titlespacing*{\section}{0pt}{4pt}{4pt}
\titlespacing*{\subsection}{0pt}{15pt}{1pt}
%%%END OF PAGE SETUP%%%

%%%FOOTER%%%
\fancyfoot{}
\fancyfoot[LO,RE]{\vspace{-7.1pt}\includegraphics[height=9pt]{head_foot/LF}}
\fancyfoot[CO]{\vspace{-7.1pt}\hspace{13.2cm}\includegraphics{head_foot/RF}}
\fancyfoot[CE]{\vspace{-7.2pt}\hspace{-14.2cm}\includegraphics{head_foot/RF}}
\fancyfoot[RO]{\footnotesize{\sffamily{1--\pageref{LastPage} ~\textbar  \hspace{2pt}\thepage}}}
\fancyfoot[LE]{\footnotesize{\sffamily{\thepage~\textbar\hspace{3.45cm} 1--\pageref{LastPage}}}}
\fancyhead{}
\renewcommand{\headrulewidth}{0pt} 
\renewcommand{\footrulewidth}{0pt}
\setlength{\arrayrulewidth}{1pt}
\setlength{\columnsep}{6.5mm}
\setlength\bibsep{1pt}
%%%END OF FOOTER%%%

%%%FIGURE SETUP - please do not change any commands within this section%%%
\makeatletter 
\newlength{\figrulesep} 
\setlength{\figrulesep}{0.5\textfloatsep} 

\newcommand{\topfigrule}{\vspace*{-1pt}% 
\noindent{\color{cream}\rule[-\figrulesep]{\columnwidth}{1.5pt}} }

\newcommand{\botfigrule}{\vspace*{-2pt}% 
\noindent{\color{cream}\rule[\figrulesep]{\columnwidth}{1.5pt}} }

\newcommand{\dblfigrule}{\vspace*{-1pt}% 
\noindent{\color{cream}\rule[-\figrulesep]{\textwidth}{1.5pt}} }

\makeatother
%%%END OF FIGURE SETUP%%%

%%%TITLE, AUTHORS AND ABSTRACT%%%
\twocolumn[
  \begin{@twocolumnfalse}
\vspace{3cm}
\sffamily
\begin{tabular}{m{4.5cm} p{13.5cm} }

 & \noindent\LARGE{\textbf{Slow and Long-ranged Dynamical Heterogeneities in Dissipative Fluids}} \\%Article title goes here instead of the text "This is the title"
\vspace{0.3cm} & \vspace{0.3cm} \\

 & \noindent\large{Karina E. Avila,\textit{$^{a,b}$} Horacio
  E. Castillo,\textit{$^{b}$} Katharina Vollmayr-Lee,\textit{$^{c}$}
  and Annette Zippelius\textit{$^{a}$}} \\%Author names go here instead of "Full name", etc.

& \noindent\normalsize{
A two-dimensional bidisperse   %submit2nd vs
granular fluid is shown to exhibit
pronounced long-ranged dynamical heterogeneities as dynamical arrest
is approached. Here we focus on the most direct approach to study
these heterogeneities: we identify clusters of slow particles and
determine their size, $N_c$, and their radius of gyration, $R_G$.
We show that $N_c\propto R_G^{d_f}$,
providing direct evidence that the most immobile particles arrange in
fractal objects with a fractal dimension, $d_f$, that is observed to
increase with packing fraction $\phi$. The cluster size distribution
obeys scaling, approaching an algebraic decay in the limit of
structural arrest, i.e., $\phi\to\phi_c$. Alternatively, dynamical
heterogeneities are analyzed via the four-point structure factor
$S_4(q,t)$ and the dynamical susceptibility $\chi_4(t)$. $S_4(q,t)$ is
shown to obey scaling in the full range of packing fractions,
$0.6\leq\phi\leq 0.805$, and to become increasingly long-ranged as
$\phi\to\phi_c$. Finite size scaling of $\chi_4(t)$ provides a
consistency check for the previously analyzed divergences of
$\chi_4(t)\propto (\phi-\phi_c)^{-\gamma_{\chi}}$ and the correlation
length $\xi\propto (\phi-\phi_c)^{-\gamma_{\xi}}$. We check the
robustness of our results with respect to 
 our definition of mobility.
  The divergences and the scaling for $\phi\to\phi_c$ suggest a 
  non-equilibrium glass transition which seems qualitatively 
  independent of the coefficient of restitution.

} 
\\%The abstrast goes here instead of the text "The abstract should be..."

\end{tabular}

 \end{@twocolumnfalse} \vspace{0.6cm}
  ]
%%%END OF TITLE, AUTHORS AND ABSTRACT%%%

%%%FONT SETUP - please do not change any commands within this section
\renewcommand*\rmdefault{bch}\normalfont\upshape
\rmfamily
\section*{}
\vspace{-1cm}

%%%FOOTNOTES%%%

\footnotetext{\textit{$^{a}$~Institut f\"ur Theoretische Physik,
    Georg-August-Universit\"at G\"ottingen, Friedrich-Hund-Platz 1,
    D-37077 G\"ottingen, Germany. }}
\footnotetext{\textit{$^{b}$~Department of Physics and Astronomy and
    Nanoscale and Quantum Phenomena Institute, Ohio University, Athens, OH, 45701, USA. }}
\footnotetext{\textit{$^{c}$~Department of Physics and Astronomy, Bucknell University, Lewisburg, PA, 17837, USA. }}
% \footnotetext{\textit{$^{c}$~Address, Address, Town, Country. }}

%%Please use \dag to cite the ESI in the main text of the article.
%%If you article does not have ESI please remove the the \dag symbol from the title and the footnotetext below.
%\footnotetext{\dag~Electronic Supplementary Information (ESI) available: [details of any supplementary information available should be included here]. See DOI: 10.1039/b000000x/}
%%additional addresses can be cited as above using the lower-case letters, c, d, e... If all authors are from the same address, no letter is required

%% \footnotetext{\ddag~Additional footnotes to the title and authors
%% can be included \emph{e.g.}\ `Present address:' or `These authors
%% contributed equally to this work' as above using the symbols:
%% \ddag, \textsection, and \P. Please place the appropriate symbol
%% next to the author's name and include a \texttt{\textbackslash
%% footnotetext} entry in the the correct place in the list.} 

%%%END OF FOOTNOTES%%%

%%%MAIN TEXT%%%%

\section{Introduction}

%**********

Supercooled liquids, colloidal suspensions, and granular systems show
evidence of strong fluctuations as they approach dynamical
arrest. These fluctuations are associated with the presence of
cooperative dynamics, and in particular with the presence of {\em
  dynamical heterogeneity\/}: some regions are populated by more
mobile particles, and relax much faster than other regions, which
contain slower particles. Experiments and simulations agree that the
heterogeneity becomes dramatically stronger when the glass transition
is
approached~\cite{Ediger2000,Berthier2011,Dauchot2005,Russell2000,Berthier2011-2}.
This phenomenon has been observed in structural glasses as well as in
colloidal suspensions and granular materials.

  Despite similarities on a phenomenological level, it is
  still controversial, how the glass transition~\cite{Debenedetti2001}
  and the jamming transition~\cite{Majmudar2007} are
  related~\cite{Liu1998,Voigtmann2011,Ikeda2012}. Whereas early work suggested
  a unified picture, more recent studies point to two separate
  transitions~\cite{Ikeda2012}, one at finite $T$ and comparatively
  low density and the other one at $T=0$ and high density. Jamming
  without applied shear is a packing problem and hence the same for
  elastically and inelastically colliding hard particles. On the
  other hand, 
  dissipation in particle-particle collisions
  causes the
  granular fluid to be inherently out of equilibrium -- in contrast to
  a thermal glass which falls out of equilibrium at the glass
  transition. Hence it is important to understand to what degree the
  phenomenon of dynamical heterogeneity in non-equilibrium systems is
  comparable to the analogous phenomenon in ordinary structural or
  colloidal glasses. With the latter, granular fluids share the
  advantage that particle positions can be tracked over time, so that
  correlated dynamics is accessible to experiment.

We analyze data from event-driven numerical simulations of 
a homogeneously driven two-dimensional hard sphere system. This
granular system resembles experiments performed on air
tables. In those experiments~\cite{Abate2007}, air is injected into
the system in order to restore the energy that is lost to dissipation
in interparticle collisions. The systems we consider contain large
numbers of particles, between $3.6 \times 10^5$ and $4.0 \times 10^6$.
Thereby finite size effects are significantly reduced. Furthermore,
large system sizes are a prerequisite to study fluctuations on large
spatial scales which is at the heart of our study.

 We observe 
a diverging relaxation time, $\tau_{\alpha}$, of the average overlap $Q(t)$, 
which allows us to identify a critical density, $\phi_c$. 
Time-density superposition is shown to be violated, but data
for different packing fractions and coefficients of restitution can be
collapsed with an empirical scaling function. We then use several
approaches to analyze the growing range and strength of the dynamic
heterogeneities as a function of packing fraction $\phi$:

\begin{itemize}
\item{First, we identify clusters of slow and fast particles and compute the cluster size distribution as well as the average
radius of gyration $R_g$. }
\item{Second, we compute the dynamical susceptibility $\chi_4(t)$ measuring the number of correlated particles.}
\item{Third, we analyze the growing length scale of dynamical
    heterogeneity by means of the four--point structure factor
    $S_4(q,t)$, which is the correlation function of the density at
    two different points for two different times.
    $S_4(q,t)$ obeys scaling and $\xi$ can be obtained from low $q$ behavior.
  }
\item{Fourth, we use finite size scaling for $\chi_4(\tau_{\alpha})$
as a consistency check for the correlation length and the fractal dimension as obtained from the above approaches.}
\end{itemize}

These results are interpreted in terms of a glass transition in a non--equilibrium fluid with pronounced dynamical heterogeneities.
Particle displacements are
strongly heterogeneous, as signaled by overpopulated tails in the
distribution of particle displacements. The latter displays an	
exponential tail for times comparable or greater than $\tau_{\alpha}$.
Clusters of slow particles are growing in size and number. The
distribution of cluster sizes obeys scaling and approaches an
algebraic decay as $\phi\to\phi_c$. Relating the radius of gyration to
cluster size shows that clusters are fractals, which compactify as
dynamical arrest is approached. The peak of the four-point
susceptibility, $\chi_4(t)$, increases dramatically as $\phi\to\phi_c$
and simultaneously the time of occurrence of the peak increases,
comparably to $\tau_{\alpha}$. Spatial fluctuations of the overlap are
measured by $S_4(q,t)$, which obeys scaling, and allows us to extract a
dynamic correlation length $\xi(t)$. Both $\chi_4(\tau_{\alpha})$ and
$\xi(\tau_{\alpha})$ are found to diverge algebraically as
$\phi\to\phi_c$, so that we obtain another estimate $d_f'$ for the fractal
dimension by relating average cluster size to the correlation
length, $\chi_4(\tau_{\alpha})\propto\xi^{d_f'}({\tau_{\alpha}})$. The
resulting fractal dimension is 
independent of packing fraction -- presumably because we do not
resolve cluster sizes. Finite size scaling analysis for $\chi_4$ as a
function of $\phi$ and $N$ provides a consistency check for the
critical behavior of $\chi_4(\tau_{\alpha})$ and $\xi({\tau_{\alpha}})$.
A short summary of some of our results has been published
in~\cite{AvilaPRL2014}

 This paper is organized as follows. In Sec.~\ref{MS} we
  describe the system and the simulation methods. In
  Sec.~\ref{results} we discuss our results, starting in~\ref{sec:overlap} with the slowing
  down of the dynamics as quantified by the dynamic
  overlap. In Sec.~\ref{PD} we present results for the distribution of
  particle displacements at different packing fractions. In
  Sec.~\ref{cluster-sec}, we analyze clusters of slow and fast
  particles in terms of the radius of gyration $R_g$ and the cluster
  size distribution. In Secs.~\ref{chi} we analyze the dynamical
  susceptibility $\chi_4(t)$ and the correlation length $\xi(t)$ at $t=\tau_{\alpha}$ as
  functions of the packing fraction. We explore
  the dependence of $\chi_4$ and $\xi$
  on (i) the system size $N$, (ii) the time difference $t$, 
  (iii) the cutoff parameter $a$ of the overlap
  function, and (iv) the coefficient of restitution
  $\varepsilon$. We confirm that the results for $\xi$ are 
robust with respect to details of the analysis such as $a$, 
the fit range and the fitting function for $S_4(q,t)$.
Finally, in Sec.~\ref{conclusions} we discuss our conclusions.  

\section{Model and Simulation details}\label{MS}
The model consists of a 2D system of hard disks which only interact
via two-body inelastic (or elastic) collisions, without a rotational
degree of freedom. The system is composed of particles of two sizes
with a $50:50$ composition, i.e., it is {\em bidisperse}. The ratio of
particle radii is given by $r_2/r_1 \approx 1.43$, where $r_1$ denotes
the radius of the small particles and $r_2$ denotes the radius of the
large particles. This is the same system as the one presented in
Ref.~\cite{Gholami2011}, which can also be consulted for
additional details of the simulation.

The change in the velocities of two colliding particles, particle {\em
  i} and particle {\em j}, is given by
\begin{equation}\label{varepsilon}
({\bf g} \cdot {\bf n})^{'}=-\varepsilon({\bf g} \cdot {\bf n}),
\end{equation}
where ${\bf g={\bf v}_i}-{\bf v}_j$ is the relative velocity, ${\bf
  n}=({\bf r}_i-{\bf r}_j)/|{\bf r}_i-{\bf r}_j|$ is a unit vector
that connects the center of the two disks and $\varepsilon$
corresponds to the coefficient of restitution, which is constant
($\varepsilon=1$ in the elastic case).  The primed quantities refer to
post-collisional velocities while the unprimed ones refer to
pre-collisional velocities. Therefore, the velocities of the two
disks after a collision are given by
\begin{equation}
m_i{\bf v}^{'}_i = m_i{\bf v}_i-\frac{m_im_j}{m_i+m_j} (1+\varepsilon)
({\bf g}\cdot{\bf n}){\bf n} 
\end{equation}
and
\begin{equation}
m_j{\bf v}^{'}_j = m_j{\bf v}_j+\frac{m_im_j}{m_i+m_j} (1+\varepsilon)
({\bf g}\cdot{\bf n}){\bf n}, 
\end{equation}
where $m_i$ corresponds to the mass of particle $i$. In our
simulations, constant mass density is assumed for all particles,
therefore the mass ratio of the particles is given by $m_2/m_1 =
(r_2/r_1)^2$.

In these systems, the driving of the particles is important to
compensate the energy dissipation due to collisions. In experiments,
this driving can be done by various methods, for example by shearing
the boundaries~\cite{Dauchot2005}, by applying frictional forces
through a moving surface that is in contact with the
particles~\cite{Lechenault2008, Wortel2014}, or by blowing an air
current through the system~\cite{Keys2007}. In our simulations, the
energy is fed homogeneously by bulk driving, in a way that is
comparable to the bulk driving in~\cite{Keys2007}. Energy is fed to
the system by applying instantaneous ``kicks'' to randomly chosen
pairs of particles. For each particle in the pair, the velocity
changes according to
\begin{equation}
m_i {\bf v}^{'}_i(t)= m_i {\bf v}_i(t)+p_{Dr}{\bf \kappa}_i(t),
 \end{equation}
 where $p_{Dr}$ is the driving amplitude and ${\bf \kappa}_i(t)$ is a
 Gaussian random vector giving the direction of the driving. The two
 particles are given opposite momenta of equal magnitude to ensure
 conservation of the total momentum of the system~\cite{Espanol1995}.
 The magnitude of the kick,
   $p_{Dr}^2=(1-\varepsilon^2)\frac{m_1m_2}{m_1+m_2}$, is chosen to
   vanish in the elastic limit. The driving frequency is chosen to be
   equal to the Enskog collision frequency,
   $\omega_{\text{coll}} = 2.59 \phi G_c \sqrt(1/\pi)$, where $G_c$ is
   the pair correlation at contact (for details see
   Refs.~\cite{Fiege2009,Gholami2011}). All results shown in this
 paper are presented in reduced units, where the length unit
 corresponds to $r_1$ and the mass unit correspond to $m_1$. Also, the
 time unit is set such that the kinetic energy, averaged over all the
 particles, is unity:
 $\frac{1}{2N_{\text{tot}}}\sum_i^{N_{\text{tot}}}m_i{{\bf
     v}_i}^2(t=0)=1$.

In this work, we analyze several simulation datasets. Some of the
simulations were performed on a system containing
$N_{\text{tot}}=4,000,000$ particles, for the
packing fractions $\phi=0.60$, $0.65$, $0.70$, $0.72$, $0.74$, $0.76$
and $0.78$, which were also used in Ref.~\cite{Gholami2011} in a
different analysis, and $N_{\text{tot}}=360,000$ for the higher
packing fractions $\phi=0.805$, $0.80$, $0.795$, $0.79$, $0.785$ and
$0.77$. We performed simulations for all packing fractions mentioned
above for the coefficient of restitution $\varepsilon=0.90$. Also, we
have additional simulations, for $\phi=0.79$, $0.78$, $0.76$, $0.74$
and $0.72$ for $\varepsilon=1.00$, $0.80$ and $0.70$. Moreover, for
each packing fraction $\phi$ we select for the analysis the time window such that the
system is in a steady state.

\subsection{Averages in our results}\label{avg}
Before introducing our results, this section is dedicated to
explaining the main procedures used to analyze our data and specify the
notation for spatial and temporal averages.

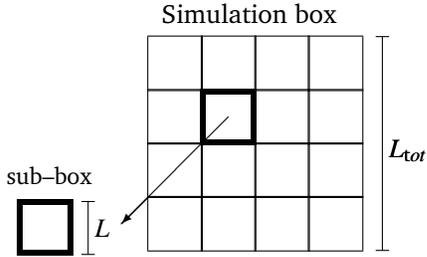
\begin{figure}[ht] 
\setlength{\unitlength}{0.14in}	% selecting unit length 
\centering	% used for centering Figure 
\begin{picture}(25,11)	% picture environment with the size (dimensions)
% 32 length units wide, and 15 units high. 
\put(10,3){\framebox(2,2)}
\put(12,3){\framebox(2,2)}
\put(14,3){\framebox(2,2)}
\put(16,3){\framebox(2,2)}
\put(10,5){\framebox(2,2)}
\put(14,5){\framebox(2,2)}
\put(16,5){\framebox(2,2)}
\put(10,7){\framebox(2,2)}
\put(12,7){\framebox(2,2)}
\put(14,7){\framebox(2,2)}
\put(16,7){\framebox(2,2)}
\put(10.5,9.5){\large Simulation box}
\put(18.75,1){\line(0,1){8}}
\put(18.5,9){\line(1,0){.5}}
\put(18.5,1){\line(1,0){.5}}
\put(19,4.5){\large $L_{\text tot}$}
\put(10,1){\framebox(2,2)}
\put(12,1){\framebox(2,2)}
\put(14,1){\framebox(2,2)}
\put(16,1){\framebox(2,2)}
\put(7.75,.85){\line(0,1){2}}
\put(7.5,2.83){\line(1,0){.5}}
\put(7.5,.85){\line(1,0){.5}}
\put(8,1.5){\large $L$}
\linethickness{0.7mm}
\put(12.15,5.15){\framebox(1.7,1.7)}
 \put(13,6){\vector(-1,-1){4}}
 \put(5.3,1){\framebox(1.7,1.7)}
\put(4.7,3.5){sub--box}
\put(19,4.5){\large $L_{\text tot}$}
\end{picture} \caption{The simulation box, of total area
  $(L_{\text{tot}})^2$, contains $N_{\text{tot}}$ particles, and is
  divided into sub--boxes of area $L^2$. Each sub-box contains a
  number $N_{{\bf r}}$ of particles, that fluctuates in different
  sub--boxes. } 
    \label{fig:box}	 
    \end{figure}

{\bf Sub--box analysis}: When probing spatial heterogeneities, we need
to compute fluctuations of two-point correlations, such as
fluctuations of the incoherent van Hove function. 
In order to do so, we divide our simulation box of total area
$(L_{\text{tot}})^2$, which contains the $N_{\text{tot}}$ particles,
into sub-boxes $B_{\bf{r}}$ of equal area $L^2$, as shown in Fig.~\ref{fig:box}. We
select $L$ such that the sub--boxes accommodate, on average, a desired
number of particles $N$, depending on the analysis performed. The
choice of $N$ will be specified in each case. However, even when $N$
is fixed, the number of particles per sub--box, $N_{{\bf r}}$, and the
particle concentration of big and small particles in general vary
between different sub--boxes. The notation $N_{{\bf r}}$ denotes the
number of particles of a sub--box centered at a point ${\bf r}$.

Notice that the total number of sub--boxes can be calculated as the
ratio $N_{\text{tot}}/N$. In general, the more particles $N_{\text
  {tot}}$, the better the statistics in the analysis.
The spatial average over the whole sample is denoted by 
$\langle \cdots \rangle$.

{\bf Time average}. Besides space averaging,
we use time averaging for some calculations to improve the statistics,
especially for the packing fractions $\phi=0.77$, $0.785$, $0.79$,
$0.795$, $0.80$ and $0.805$, for which the system contains fewer
particles than for the rest of the packing fractions. Time averages are denoted by  $\overline{\cdots}$.
It is worth
noting that time averaging is possible because the simulations are
done in the stationary state, i.e., the system is not aging. This
means that an average over different choices of the starting time $t_0$ can be performed
for quantities that have a dependence on the time difference.

\section{Results}\label{results}

\subsection{Overlap}\label{sec:overlap}

Before we examine dynamic heterogeneities, i.e., four-point correlations,
let us first look at the relaxation of the system and the relevant 
time-scales associated with this relaxation. We begin with the
two-point correlation, the overlap
\begin{equation}
Q_{\bf r}(t;t_0)= \frac{1}{N}\sum_{i=1}^{N_{{\bf r}}} \theta(a-|{\bf
  r}_i(t_0+t)-{\bf r}_i(t_0)|),
\label{Q(t)-eq}
\end{equation} 
where ${\bf r}_i(t)$ is the position of particle $i$ at time $t$ and
$\theta(x)$ is the Heaviside theta function, $\theta(x) = 1$ for $x>0$
and $\theta(x) = 0$ for $x<0$. The sum runs over particles
$i=1,\ldots, N_{\bf r}$ which are at time $t_0$ in a sub--box
$B_{\bf r}$ of size $L^2$ centered at ${\bf r}$.  Here $N_{\bf r}$ is
the actual number of particles in the region at time $t_0$ and
$N \equiv \langle N_{\bf r} \rangle $ is the average number of
particles for a region of the given size.  We refer to those particles
that moved less than a given {\em cutoff distance\/} $a$ over the time
interval between $t_0$ and $t_0+t$ as {\em slow particles\/}. The
overlap $Q_{\bf r}(t;t_0)$ is therefore the ratio between the number
of slow particles in the sub--box centered at ${\bf r}$ and the
average number of particles in the region, and hence a fluctuating
quantity.  We will study its fluctuations in
 Sec.~\ref{S4xi-sec}, but first discuss its average to identify the
  relevant timescale and discuss the relaxation to a stationary state.

\begin{figure}[b]
 \includegraphics[scale=.64]{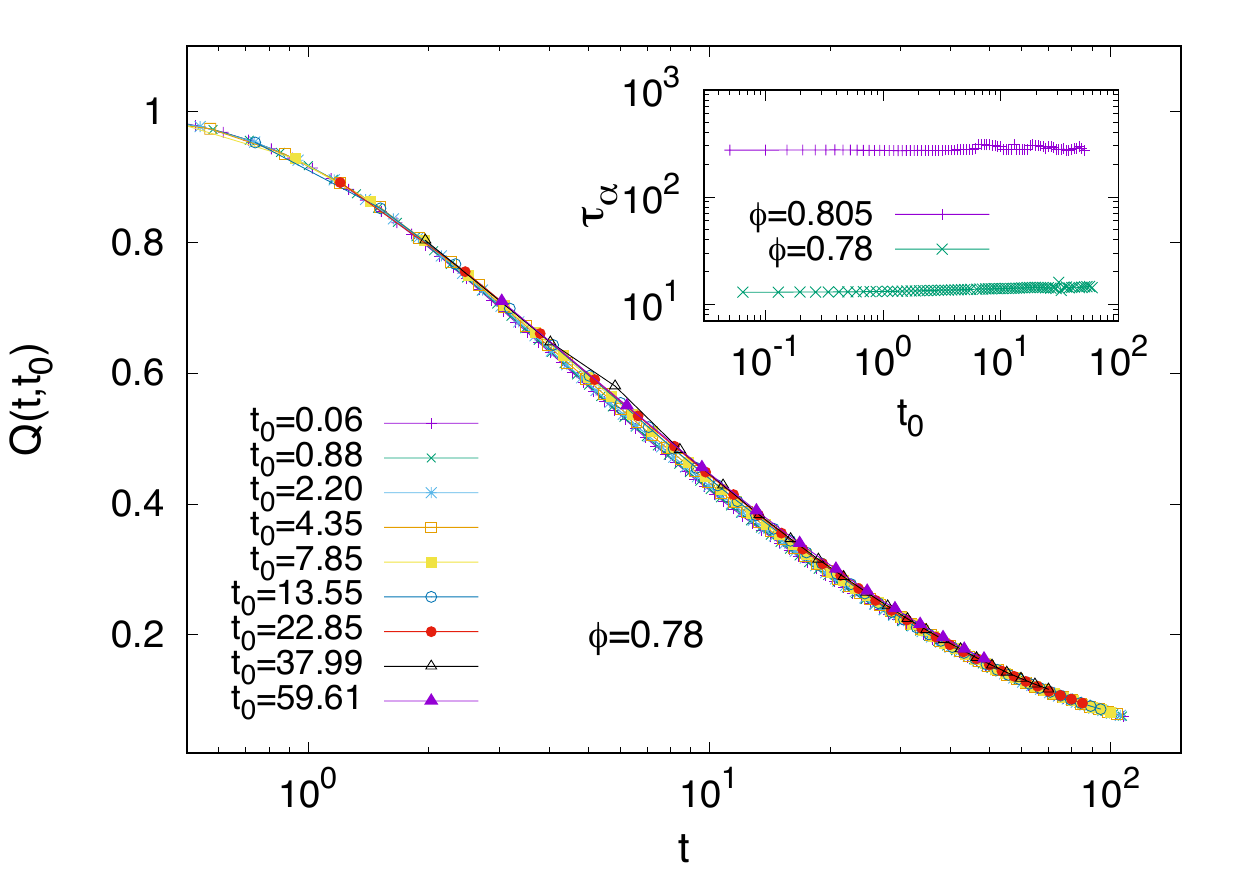}
 \caption{Overlap, $Q(t;t_0)$, for
 $\phi=0.78$ and several waiting times $t_0$; inset: relaxation time 
$\tau_{\alpha}$, defined by $Q(\tau_{\alpha},t_0)=1/e$ as a function of $t_0$ for two packing fractions $\phi=0.78$, and $0.805$.}
  \label{Q-t-t_0}
  \vspace{-0.5cm}
\end{figure}

  We want to make sure that the system has relaxed to a stationary
  state before taking measurements. The required relaxation time,
  $\tau_{\text{stat}}$, is comparable to $\tau_{\alpha}$, defined by
  $Q_r(\tau_{\alpha},t_0)=1/e$. Typically we take
  $\tau_{\text{stat}}\geq 9\tau_{\alpha}$. To check that this is
  indeed sufficient we consider the average over subboxes
\begin{equation}
Q(t;t_0) = \left<Q_{\bf r}(t;t_0)\right>,  
\label{Q_avg(t;t_0)}
\end{equation}  
which in general still depends on two times: the waiting time $t_0$
which elapses after relaxation and before taking measurements and the
time difference $t$.  In the stationary state this function should not
depend on $t_0$ anymore.  In Fig.~\ref{Q-t-t_0} we show $Q(t;t_0)$ for
packing fraction $\phi=0.78$ and many waiting times $t_0$.
The function $Q(t;t_0)$ is indeed independent of $t_0$, in other words
no aging is observed. We have checked this for all densities and show
$\tau_{\alpha}$ as a function of $t_0$ in the inset for $\phi=0.78$
and $0.805$.

\begin{figure}
 \includegraphics[scale=.64]{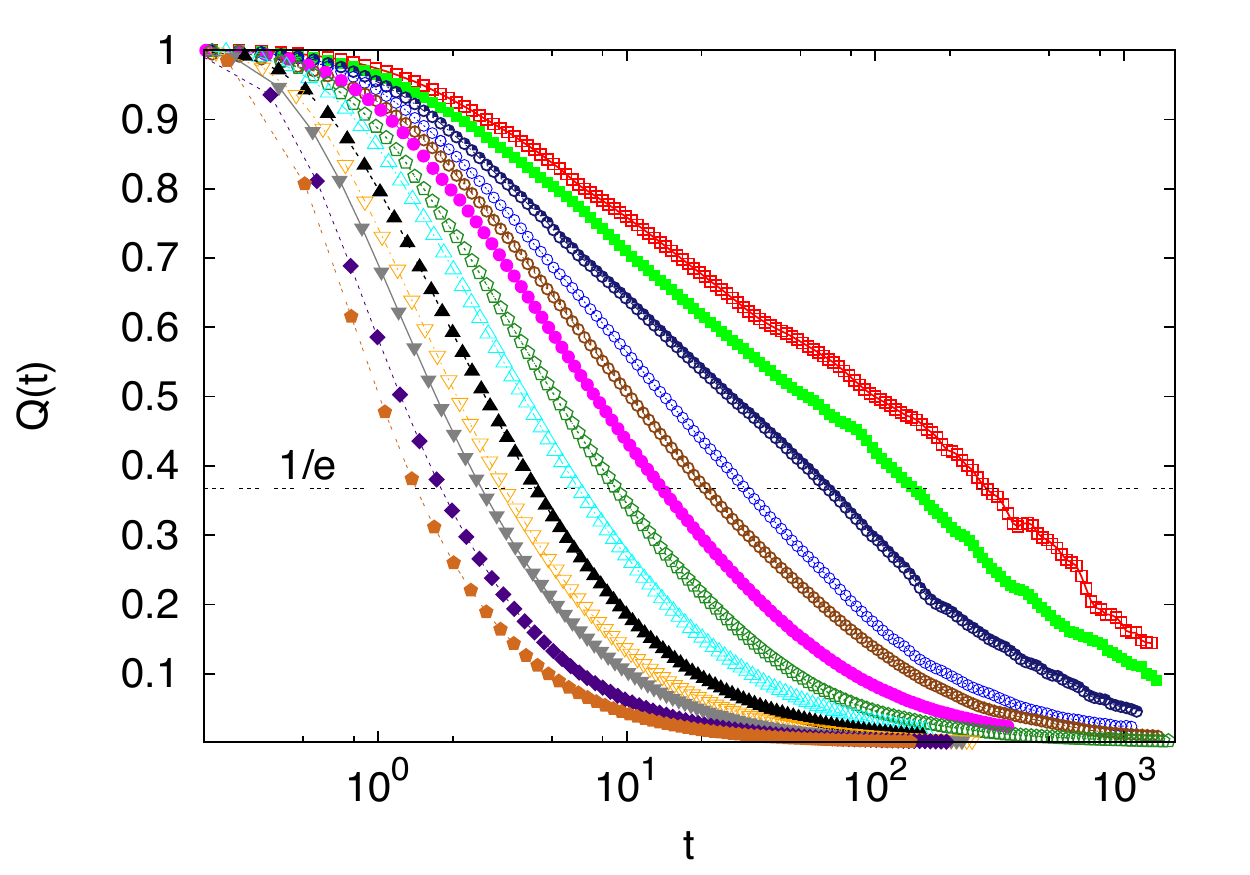}
 \caption{Slowdown of the overlap function, $Q(t)$, for
   packing fractions (left to right) $\phi=0.60, 0.65, 0.70, 0.72$, $0.74, 0.76,
   0.77$, $0.78, 0.785, 0.79, 0.795, 0.80, 0.805$; the horizontal line
   corresponds to the value of the correlation function,
   $Q(\tau_{\alpha})=1/e$, defining $\tau_{\alpha}$. }
  \label{Q-t}
  \vspace{-0.5cm}
\end{figure}

\begin{figure}
  \begin{center}
    \includegraphics[scale=.65]{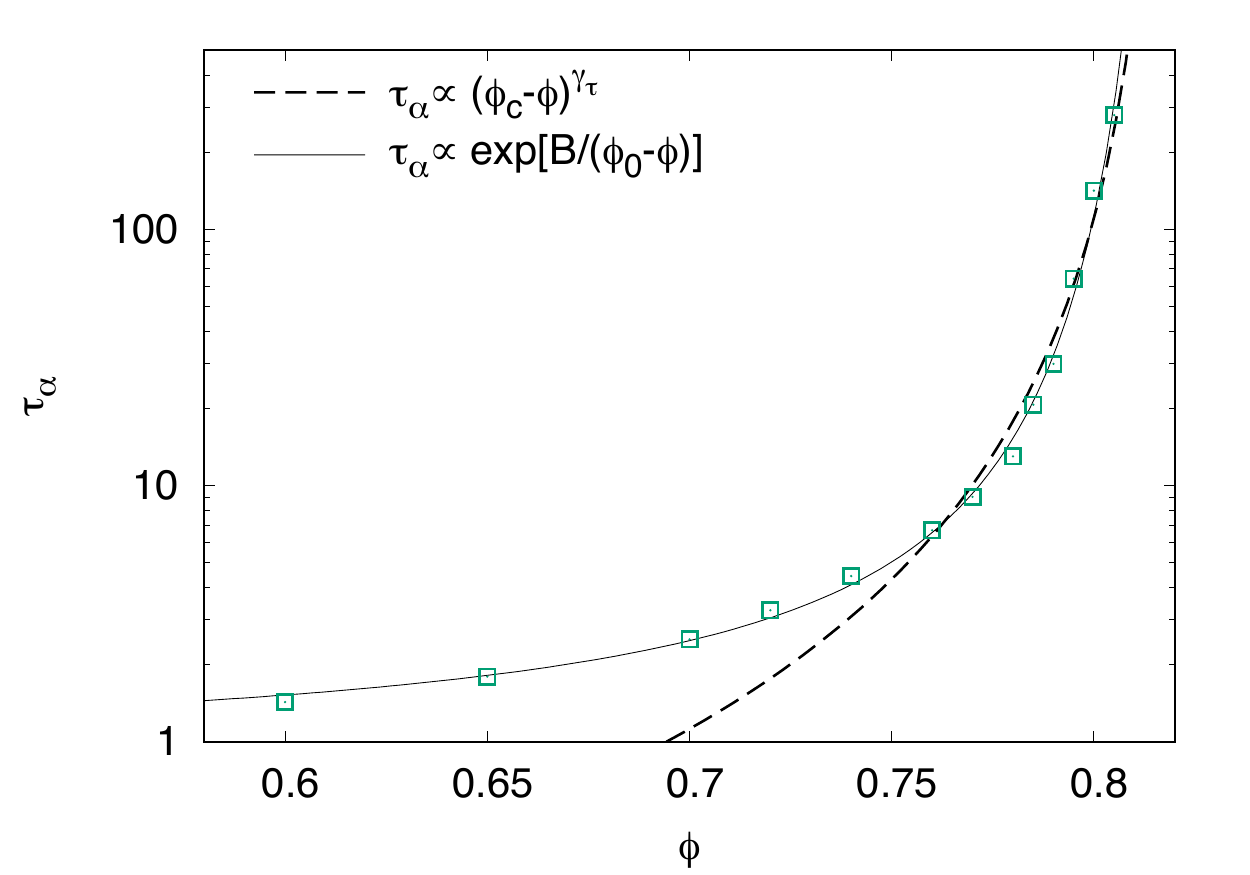} 
  \end{center}
  \caption{
    Dependence of the relaxation time on $\phi$.
    Two possible fits,
    $\tau_{\alpha}\propto (\phi_c-\phi)^{-\gamma_{\tau}}$ (dashed
    line) and $\tau_{\alpha}\propto \exp{[B/(\phi_0-\phi)]}$ (solid
    line), are shown.    } 
  \label{taueps09}
\end{figure}

Given that we are in a stationary state, we average the overlap over time 
\begin{equation}
Q(t) = \overline{\left<Q_{\bf r}(t;t_0)\right>},  
\label{Q_avg(t)}
\end{equation} 
which is shown in Fig~\ref{Q-t}. (We always choose  $a=0.6$ and $\varepsilon = 0.9$, unless stated otherwise).
%-------------
%\medskip 
In contrast to three-dimensional systems, there is almost
  no two-step relaxation~\cite{Gholami2011,FlennerSzamel2015}, instead
  the
decay of $Q(t)$ just slows down progressively as the packing fraction $\phi$ is increased. To quantify this slowdown, 
we show in Fig.~\ref{taueps09} the relaxation time
$\tau_{\alpha}$ as a function of $\phi$, where $\tau_{\alpha}$ is
defined by $Q(\tau_{\alpha})={\rm 1/e}$. 
Included in
Fig.~\ref{taueps09} are two fits: one to an inverse power law
$\tau_{\alpha}\propto (\phi_c-\phi)^{-\gamma_{\tau}}$ (dashed line) as
predicted by mode coupling theory~\cite{Bengtzelius1984,Kranz2010}, and another
to an exponential form $\tau_{\alpha}\propto \exp{[B/(\phi_0-\phi)]}$
(solid line). Both fitting forms extrapolate to a divergence of
$\tau_{\alpha}$, located at $\phi \to \phi_c=0.82$ and at $\phi \to
\phi_0=0.83$ respectively.  The exponential fit describes the curve
better across the whole $\phi$ range than the inverse power law, which
only works for a narrower range of values of $\phi$. This behavior has
been also found in other systems (see for example~\cite{Brambilla2009,Berthier2009}).

\begin{figure}
  \begin{center}
    \includegraphics[scale=.66]{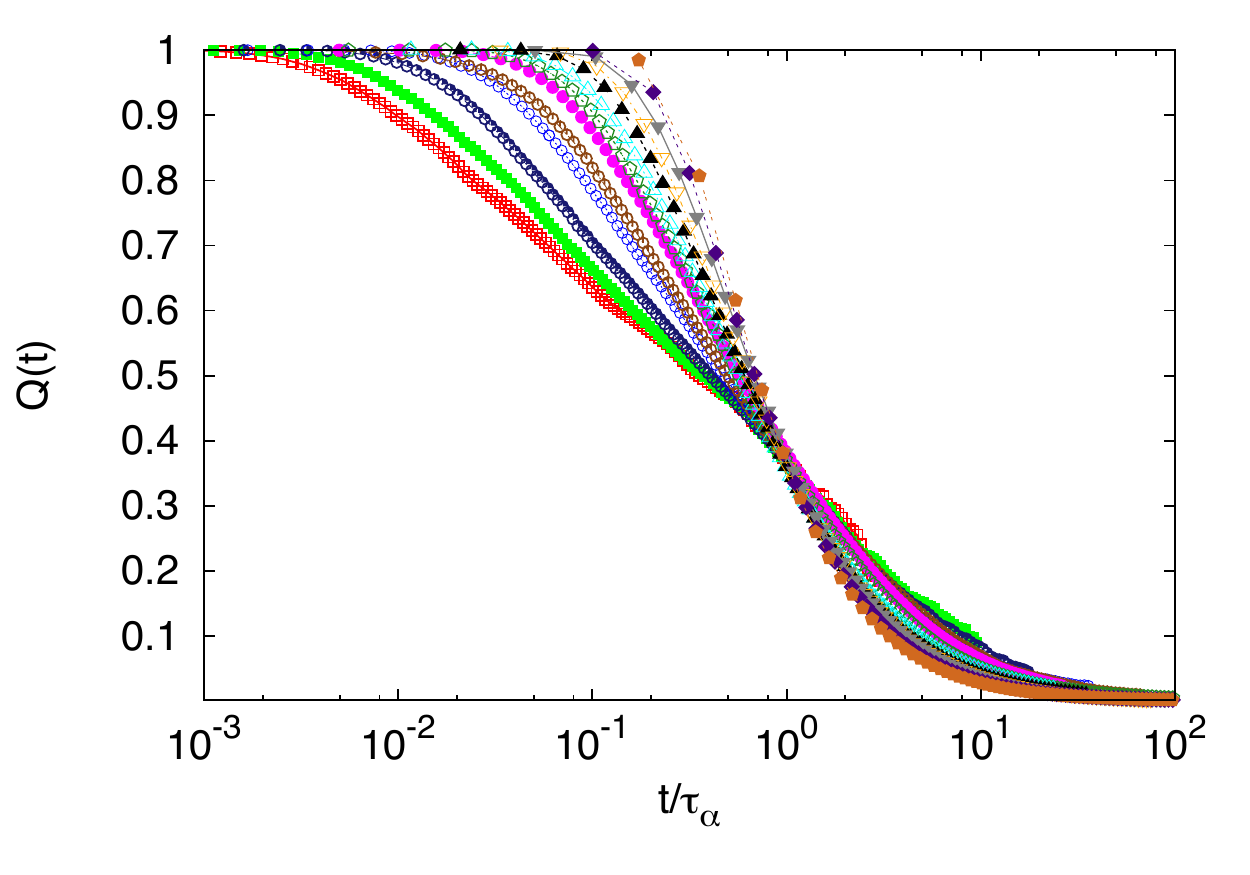} 
  \end{center}
  \caption{
    The overlap $Q$ as in Fig.~\ref{Q-t} but here as a function of
    $t/\tau_{\alpha}$ to demonstrate the breakdown of time-density
    superposition (color coding as in Fig.~\ref{Q-t}).  }
  \label{Qoftovertaualpha}
\end{figure}

\medskip Thus, our results so far are similar to previous work on
glass dynamics. Unexpected differences, however, occur in the shape
of the $Q(t)$ decay. In Fig.~\ref{Qoftovertaualpha} we test
time-density superposition, i.e., whether $Q(t)$ for different $\phi$
is scaling with $t/\tau_{\alpha}(\phi)$. Clearly, time-density
superposition does not give rise to a good data collapse for granular fluids in 2D. Similarly, in~\cite{FlennerSzamel2015}
it was found that for a 2D non--dissipative glassy system time--temperature
superposition fails, while in a similar 3D system it holds. 

\begin{figure}
  \begin{center}
    \includegraphics[scale=.665]{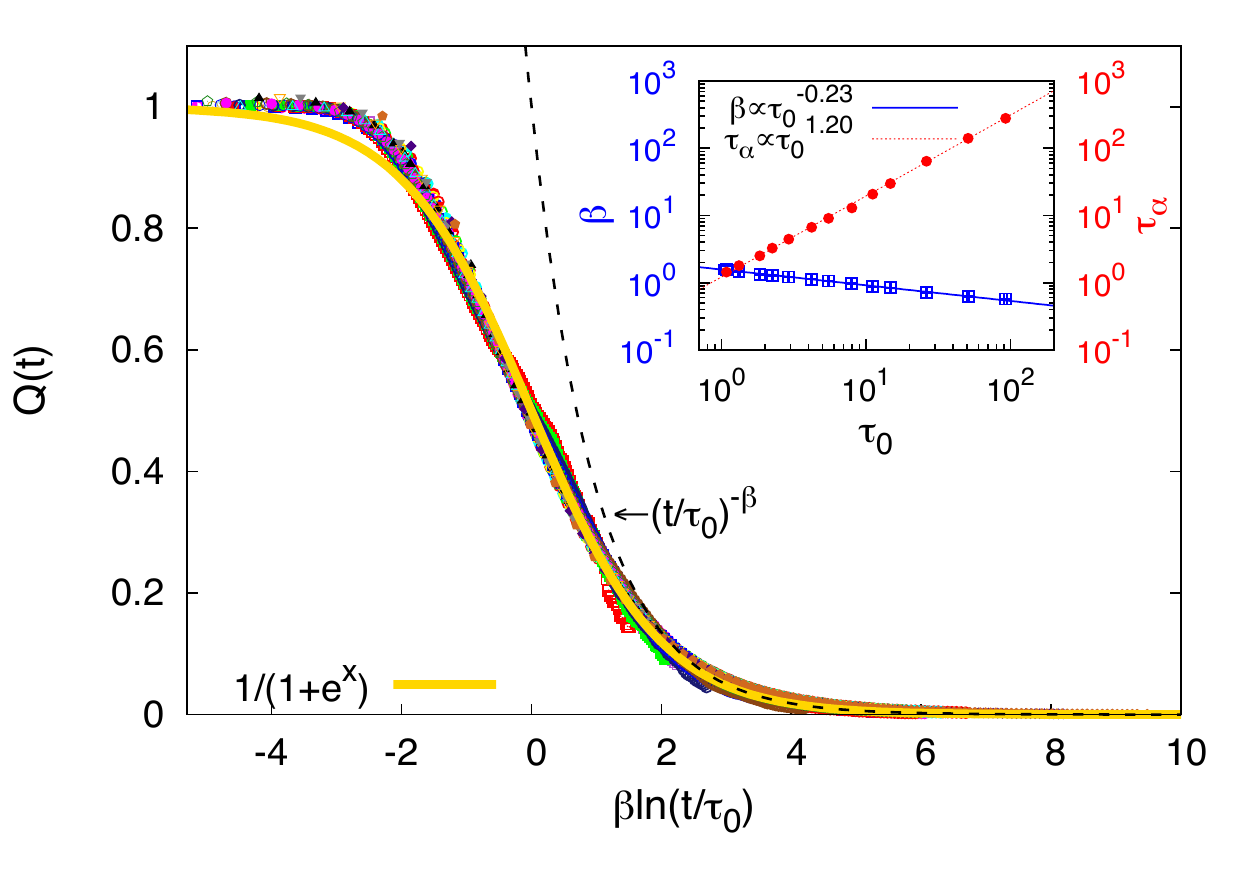} 
  \end{center}
  \caption{$Q(t)$ fitted to Eq.~(\ref{eq:Fermi-fit}) with fit
    parameters $\tau_0(\phi)$ and $\beta(\phi)$, for all simulated
    values of $\phi$ and $\varepsilon$; dashed line indicates the
    algebraic time dependence, $(t/\tau_0)^{-\beta}$ for
    $t>\tau_{\alpha}$. The inset shows that $\beta(\tau_0)$ and
    $\tau_{\alpha}(\tau_0)$ follow power laws for $\varepsilon=0.90$.
  }
  \label{Qofbetalntovertau0}
\end{figure}

\medskip
Despite the absence of time-density superposition, a
simple description of the relaxation functions is possible. 
We have fitted $Q(t)$ with the empirical form
\begin{equation}
  Q(t) = \frac{1}{\exp[\beta\ln(t/\tau_0)] + 1} =
  \frac{1}{\left({t}/{\tau_0}\right)^{\beta} + 1},
\label{eq:Fermi-fit}
\end{equation}
where the exponent $\beta$ and the characteristic time $\tau_0$ are
fitting parameters that depend on $\varepsilon$ and
$\phi$. Fig.~\ref{Qofbetalntovertau0} is a scaling plot of $Q(t)$ as a
function of $x \equiv \beta\ln(t/\tau_0)$, for all simulated values of
packing fraction
$\phi$ and coefficients of restitution $\varepsilon$ (see below for a discussion of variations of $\varepsilon$). The numerical results for different densities and restitution
coefficients show a remarkably good collapse.
Both, $\beta$ as well as $\tau_0$, follow a power law as a function of
$\tau_{\alpha}$ (see inset of Fig.~\ref{Qofbetalntovertau0}).

The collapse of the data to the empirical scaling function
Eq.~(\ref{eq:Fermi-fit}), implies an algebraic decay $\propto (t/\tau_0)^{-\beta}$
for times $t\geq \tau_{\alpha}$ as indicated in
Fig.~\ref{Qofbetalntovertau0}.

\subsection{Heterogeneous Particle Displacements}\label{PD}

Next we investigate the heterogeneity in the particle displacement. 
Whereas in the following section we will study the {\rm spatial} 
distribution of ``fast'' and ``slow'' particles, we quantify 
in this section the disparity between fast and slow particles.

\begin{figure}[b!]
  \begin{center}
    \includegraphics[scale=.66]{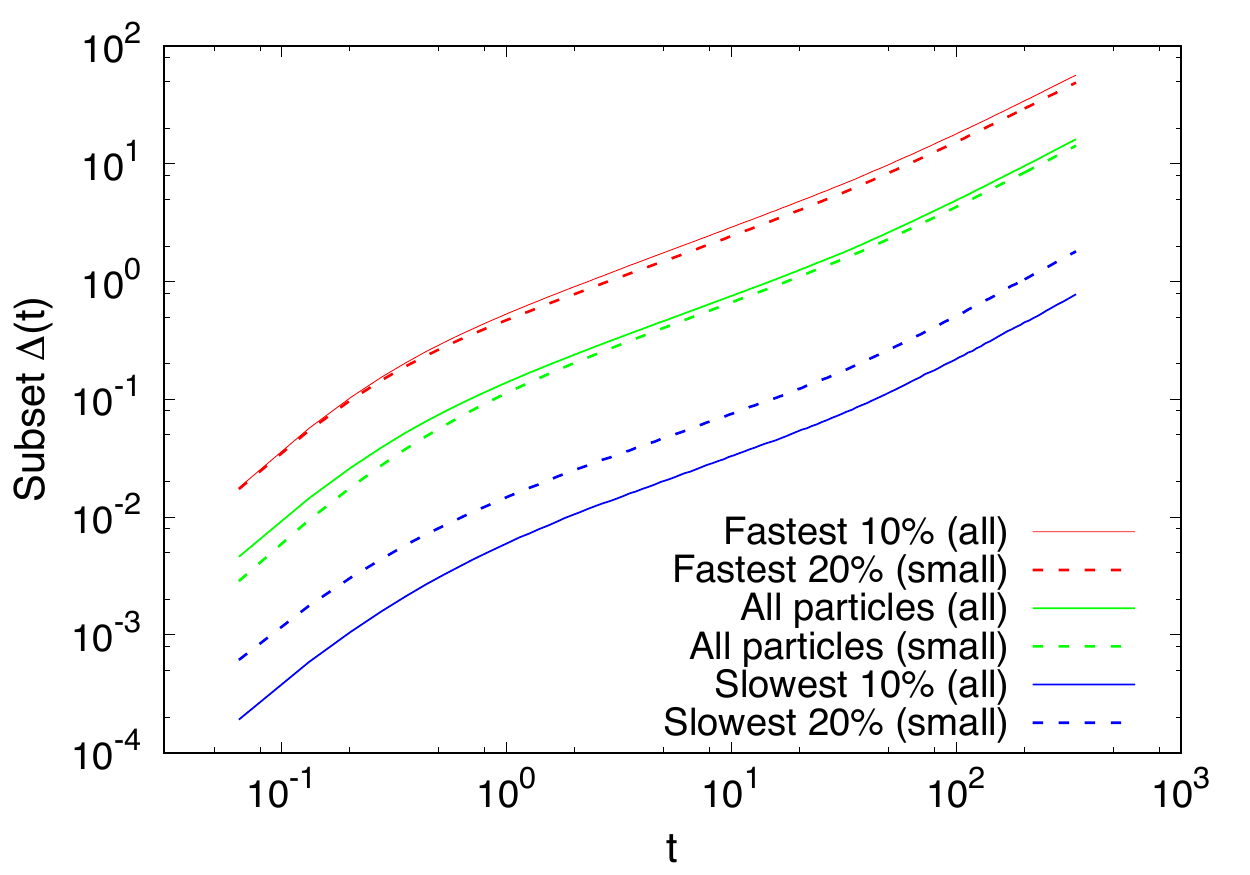}
  \end{center}
  \caption{MSD for different subsets of particles,
    either restricted to small particles (dashed lines) or for all
    particles (solid lines) for packing fraction $\phi=0.78$; the
    upper curves corresponds to the fastest particles, the middle
    curves to all particles  and the lower ones to the slowest
    particles. 
  } 
  \label{MSD}
\end{figure}

\begin{figure}[ht]
  \begin{center}
    \begin{picture}(220,320)
  \put(0,150){\label{fig:dx_a} \includegraphics[scale=.64]{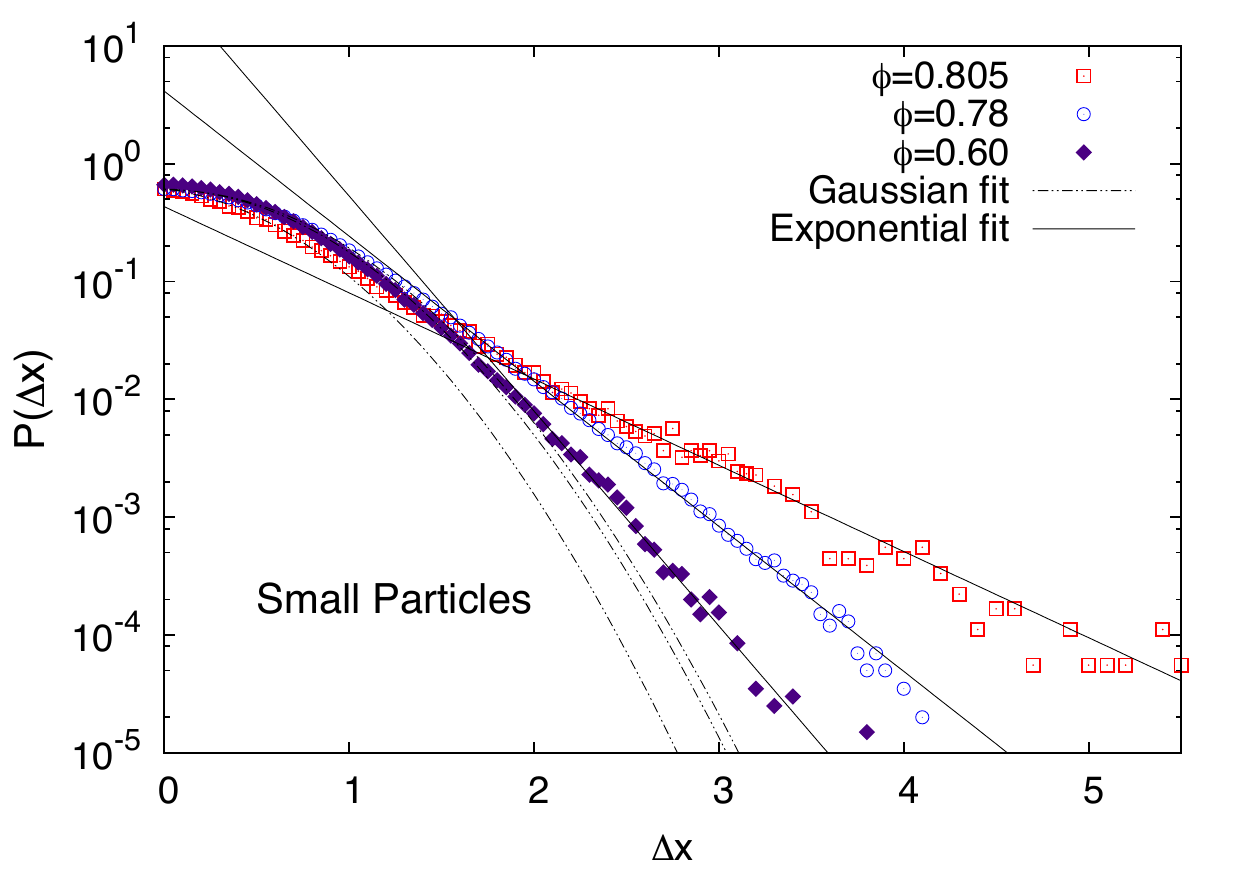}}
 \put(0,-10){\label{fig:dx_b} \includegraphics[scale=.64]{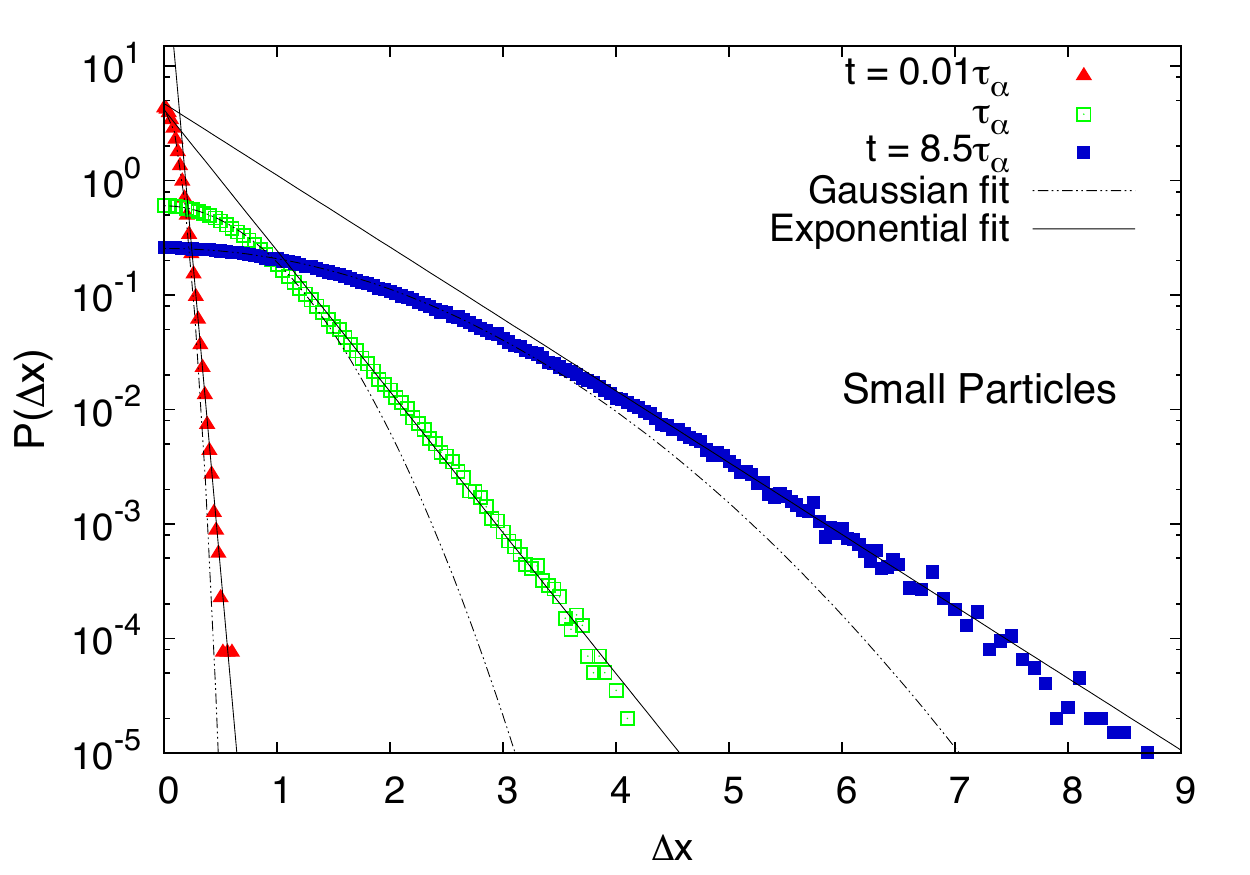}} 
 \put(52,294){\large{(a)}}
  \put(52,128){\large{(b)}}
 \end{picture}
  \end{center}
  \caption{(a) Distribution of small particle
    displacements for different packing fractions $\phi=0.805$,
    $0.78$, and $0.60$ (from right to left) at time
    $\tau_{\alpha}$. (b)  Distribution of small particle displacements
    for $\phi=0.78$ at different times $t=0.01\tau_{\alpha}$,
    $t=\tau_{\alpha}$ and $t=8.5\tau_{\alpha}$ (from left to
    right). The tails of the distributions are better described by an
    exponential fit (solid lines) than by a gaussian fit
    (dotted-dashed lines).} 
  \label{tails}
\end{figure} 

\begin{figure}[ht]
  \begin{center}
    %\subfloat{\label{fig:diffusion-coeff}  
    \label{fig:diffusion-coeff}  
\includegraphics[scale=.66]{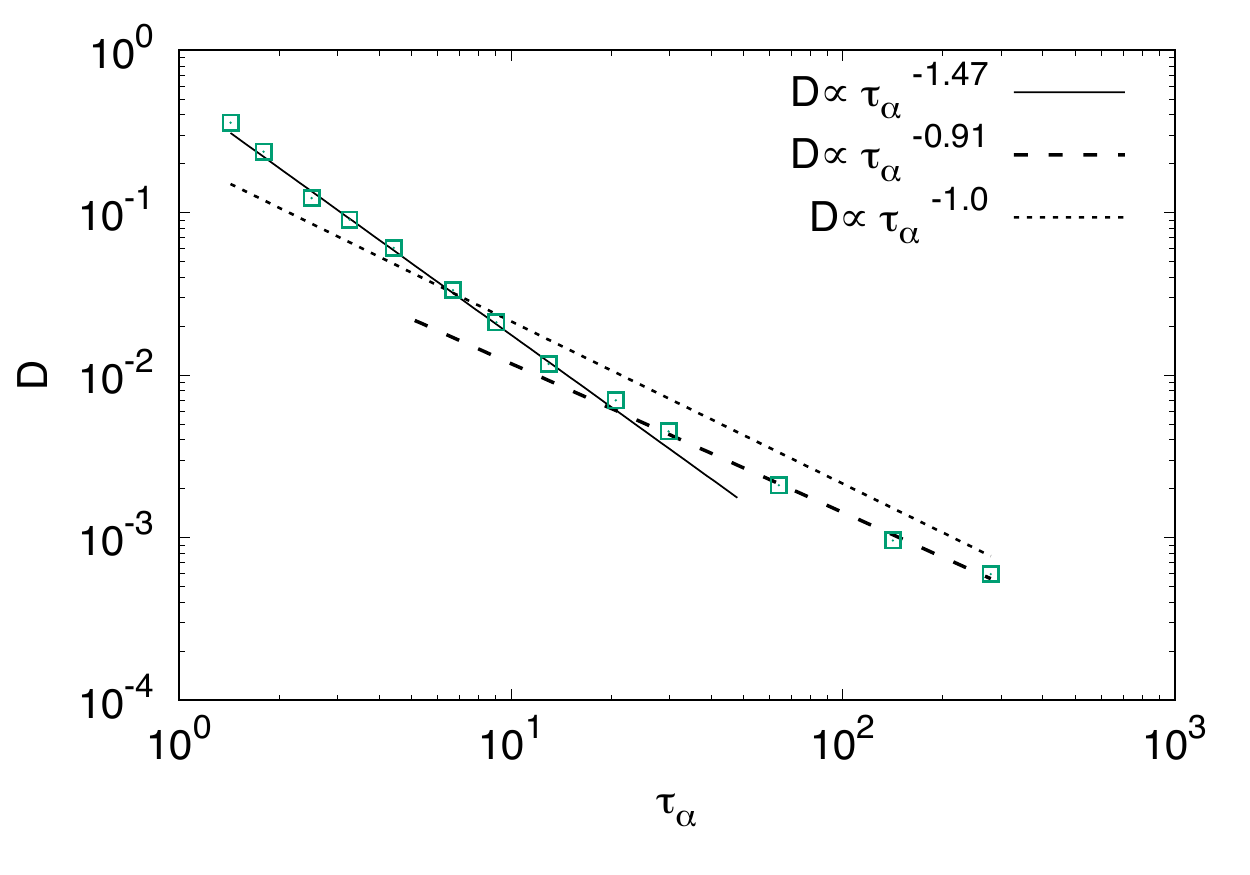}
  \end{center}
  \caption{Diffusion coefficient $D$ as a function of $\tau_{\alpha}$
    for different values of $\phi$, demonstrating the breakdown of the
    Stokes-Einstein relation as shown as a dotted line.  Instead we
    observe a crossover from $D\propto {\tau_{\alpha}}^{-\theta}$ at
    intermediate densities (solid line) to
    $D\propto {\tau_{\alpha}}^{-\theta'}$ at the highest densities
    (dashed line).}
  \label{D-vs-tau}
\end{figure} 

Fig.~\ref{MSD} shows the mean square displacement (MSD) for all 
particles (green solid line) and for small particles (green dashed--dotted line)
given by 
\begin{equation} 
  \Delta(t)=\frac{1}{N_{\text{set}}}\sum_{i=1}^{N_{\text{set}}} 
        ({\bf r}_i(t_0+t)-{\bf r}_i(t_0))^2, 
 \label{MSD-eq} 
\end{equation}
where $N_{\text{set}}$ is the total number of particles belonging
to the subset considered, i.e., $N_\text{set}=N_\text{tot}$ for all particles (see Sec.~\ref{avg} for the definition of
$N_{\text{tot}}$) and ${\bf r}_i(t)$ is the position of particle {\em i}
at time $t$. 
We furthermore show in Fig.~\ref{MSD} the MSD as a function of time $t$
for subsets defined based on the instantaneous particles` MSD 
%$({\bf r}_i(t_0+t)-{\bf r}_i(t_0))^2$:

- the fastest $10\%$ of all particles (red full line),

- the fastest $20\%$ of small particles (red dashed line),

- the slowest $10\%$ of all particles (blue full line), 

- the slowest $20\%$ of small particles (blue dashed line)

and compare them to the corresponding quantities for all
particles. Note that the sets of the slowest (fastest)
  $10\%$ refer here to an instant of time $t$ or rather a small
  interval $[t,t+\Delta t]$. For each small time interval $\Delta t$
  the selected subset is in general different. Therefore, many of the particles which are among the slowest $10\%$ for a certain time interval will not be among the slowest $10\%$ for other (shorter or longer) time intervals.

We find, in accordance with Fig.~13 of~\cite{Gholami2011}, a vast
difference between fast and slow particles. While the 10\% fastest
particles move a distance several times the radius $r_1$ of the small
particles, the slowest 10\% of particles barely move. The restriction
to small particles does not significantly change this
observation. The drastic differences of particle mobility give us an
idea of the strength of the dynamical heterogeneity.

\medskip
The full distribution of displacements in the $x$-direction, 
$\Delta x(t)=[x_i(t_0+t)-x_i(t_0)]$ at a fixed time difference $t$ is shown in 
Fig.~\ref{tails}(a) for various packing fractions and 
in Fig.~\ref{tails}(b) for several times $t$.
For a fluid far from dynamical arrest, the particles are expected to
perform a simple random walk, and the displacement distributions are
expected to have a Gaussian form $P_g(\Delta x,t)=\left(1/\sqrt{4\pi Dt}\right )
\exp[-(\Delta x)^2/(4Dt)]$, where $D$ is the diffusion
coefficient. Gaussian fits to the data, with $D$ used as a fitting
parameter, are shown with dotted-dashed lines in Fig.~\ref{tails}. We
observe that the distributions deviate strongly from the Gaussian fit
for all packing fractions and times. The tails of the distributions
follow approximately exponential behavior, $P_e(\Delta x,t)\propto
\exp(-|(\Delta x)/x_0(t)|)$, shown as solid lines. Also, the tails
become wider both for increased packing fraction and for longer
times. Exponential tails have been studied also in non-dissipative 
glassy systems~\cite{Weeks2000,Parsaeian2009,Chaudhuri} and have been
established as an indirect signature of spatial dynamical
heterogeneity~\cite{Chaudhuri}.

\medskip
Another expected consequence of the presence of heterogeneous
dynamics is that supercooled liquids near the glass transition in 3D
violate both the Stokes-Einstein relation $D\eta/T =
const$~\cite{Ediger2000,Berthier2011} connecting the diffusion
coefficient $D$ with the viscosity $\eta$, and the related condition
$D\tau_{\alpha}/T = const'$ connecting $D$ with the $\alpha$-relaxation
time $\tau_{\alpha}$. Both ratios, $D\eta/T$ and $D\tau_{\alpha}/T$,
show strong increases as the liquid approaches dynamical arrest. In
two dimensional thermal systems, a different phenomenology
has been found~\cite{Sengupta2013}: both ratios behave as power laws
as functions of temperature, even far from dynamical arrest, but the
exponents for the power laws show significant changes as the liquid
goes from the normal regime to the supercooled regime.

In our case, we focus on the relation between $D$ and
$\tau_{\alpha}$. We obtain the values of $D$ by fitting the long time
limit of the MSD (see Eq.~(\ref{MSD-eq})) with the form
$\Delta(t)=4Dt$. This is known to be problematic in 2D, because long
time tails of the velocity autocorrelation threaten the existence of
hydrodynamics~\cite{Gholami2011}. However, these tails are strongly suppressed in the
vicinity of the glass transition, so that the above naive definition of $D$ is presumably only weakly -- if at all -- affected. 

Fig.~\ref{D-vs-tau} is a plot
of $D$ as a function of $\tau_{\alpha}$, for all values of $\phi$.
For packing fractions not too close to dynamical arrest, a
power law behavior $D \propto {\tau_{\alpha}}^{-\theta}$ is found,
with $\theta \approx 1.47$.
\begin{figure*}[ht]
  \begin{center}
    \includegraphics[scale=0.5]{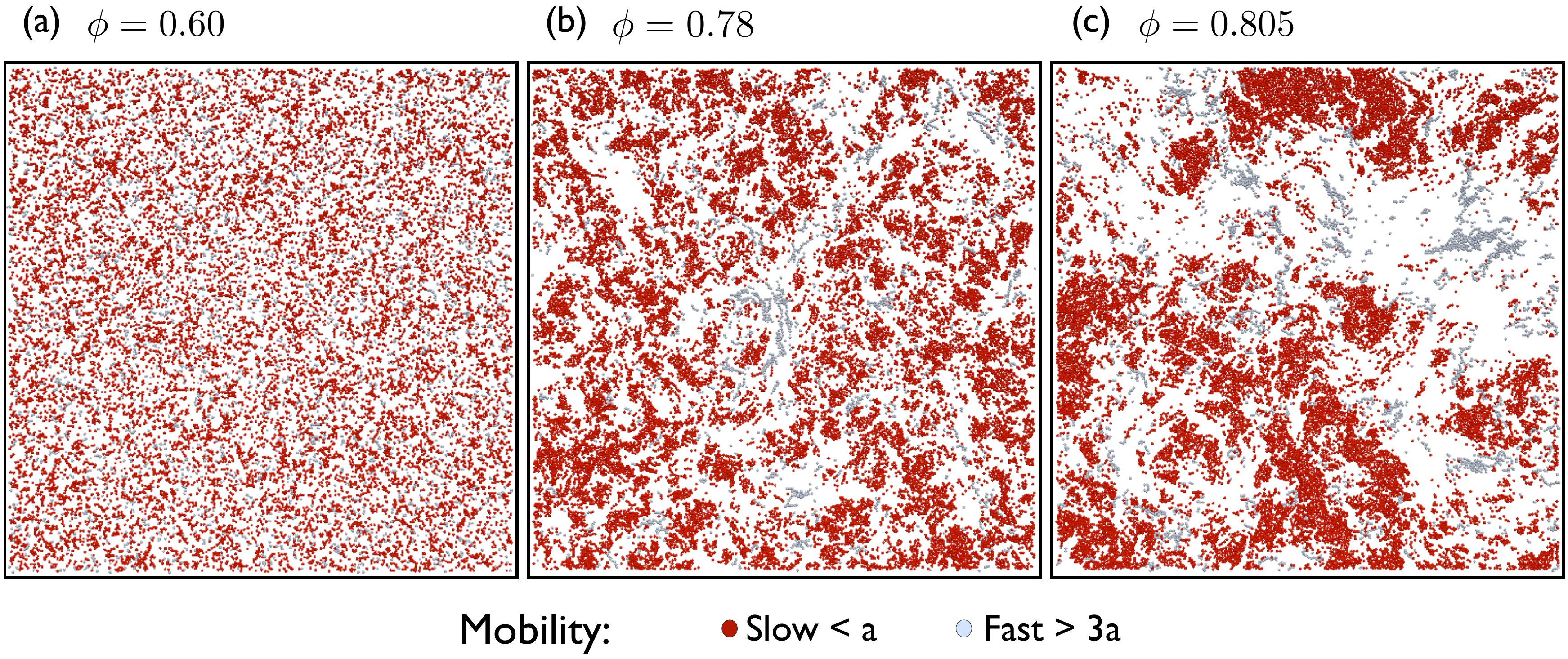}
  \end{center}
  \caption{Spatial distribution of slow and fast
    particles. Slow particles, which had a displacement shorter than
    $a$ in one relaxation time are shown in red (dark gray), fast
    particles, which had a displacement longer than $3a$ in one
    relaxation time are shown in gray (light gray). Here
    $\varepsilon=0.9$, $a=0.6$, and the different panels correspond
    to different packing fractions: (a) $\phi=0.60$, (b) $\phi=0.78$
    and (c) $\phi=0.805$. }
  \label{SD-2}
\end{figure*} 
For higher packing fractions, one observes a crossover to a power law
with a different exponent, $\theta' \approx 0.91$.  
  These results confirm the breakdown of the Stokes-Einstein relation
  for small $\tau_{\alpha}$, i.e., far away from the glass transition,
  which has also been observed~\cite{Sengupta2013} in 2D
  non-dissipative glass forming systems.

\subsection{Clusters of slow and fast particles}
\label{cluster-sec}

In this section we investigate {\em directly} the {\em spatial} 
distribution of dynamical heterogeneities.
We look at the whole system as one unit,
instead of dividing it into sub-boxes. 

To visually observe dynamical heterogeneity in our system we
color--code particles according to their mobility.  As in
Sec.~\ref{sec:overlap}, we define {\em slow particles\/} as those that
for a given time interval $t$ have a displacement smaller than the
cutoff $a$. Additionally, we define {\em fast particles\/} as those
that in the same time interval have a displacement larger than
$3a$. The spatial distribution of slow and fast particles is shown in
Fig.~\ref{SD-2}, for $\varepsilon = 0.9$ and three different packing
fractions, for a time interval $t=\tau_{\alpha}$ corresponding to the
$\alpha-$relaxation time. The fast particles are displayed in gray (light gray)
and the slow particles in red (dark gray).  We observe that both slow
and fast particles form clusters, and that in both cases the typical
size of the clusters increases as the packing fraction increases.

We now analyze quantitatively the size and shape of those clusters
for several packing fractions $\phi$ at time difference $\tau_{\alpha}$.
For each packing fraction we use several snapshots of the system,
generated starting from different initial condition. Unless
otherwise indicated, all results in this analysis are for
$\varepsilon=0.9$ and $a = 0.6 $.

Two slow/fast particles belong to the same cluster if they are linked
by a chain of nearest neighbor pairs of slow/fast particles. Since the
particles are distributed continuously in space, there is some
ambiguity in the definition of what constitutes a pair of nearest
neighbors. By convention, we say that two particles are nearest
neighbors if they are separated by a distance smaller than 
$r^{\rm  min}_{\alpha \beta}$, where $r^{\rm min}_{\alpha \beta}$ is the
position of the first minimum of the radial pair distribution
$g_{\alpha \beta}(r)$ for particles $\alpha,\beta \in \{1,2\}$ (see
Fig.~14 and Fig.~15 in~\cite{Gholami2011}).

\medskip
Let's consider one of the clusters in the system. In order to quantify
its size and shape, we define $N_c$ as the number of particles in the cluster
and its radius of gyration by
\begin{equation}
R_g=\left[\frac{1}{N_c}\sum_{i=1}^{N_c}| {\bf r}_i(\tau_{\alpha}) -
  {\bf R}_{CM}(\tau_{\alpha})|^2 \right]^{1/2},  
\end{equation}
where ${\bf r}_i(\tau_{\alpha})$ for $i = 1, \cdots, N_c$ are the
positions of the particles that belong to the cluster, and ${\bf
  R}_{CM}(t) \equiv N_c^{-1} \sum_{i=1}^{N_c} {\bf r}_i(t)$ is the position
of the cluster's center of mass.  

\begin{figure}[h]
  \begin{center}
    \begin{picture}(220,320)
  \put(-4,150){\label{fig:Rg-slow} \includegraphics[scale=.66]{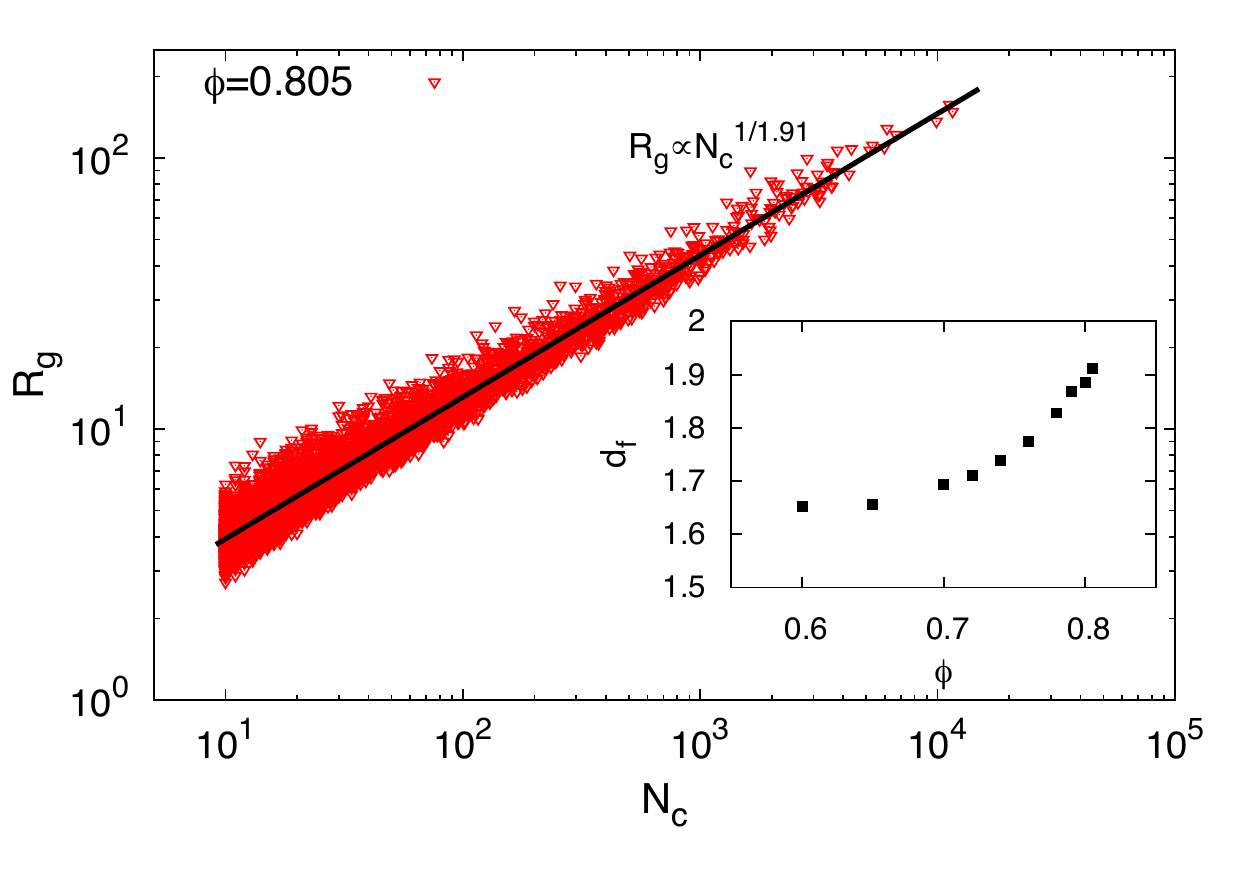}}
 \put(0,-10){\label{fig:Rg-fast} \includegraphics[scale=.64]{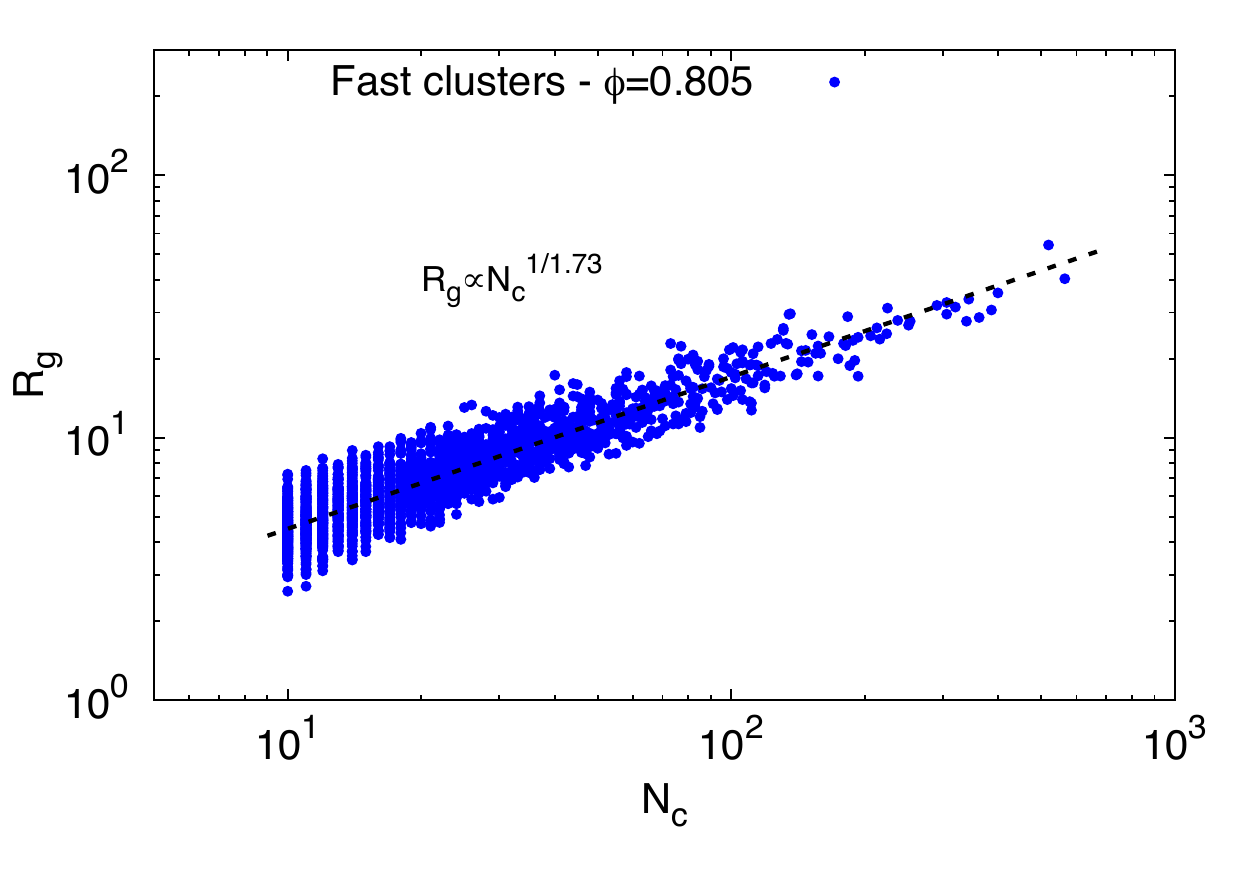}} 
 \put(40,280){\large{(a)}}
  \put(40,115){\large{(b)}}
 \end{picture}
  \end{center}
  \caption{Radius of gyration $R_g$ as a function of
    cluster size $N_c$ for $\phi=0.805$.  A power law fit of the data
    to $R_g\propto{N_c}^{1/d_f}$ gives the values $d_f\approx1.91$ for
    (a) slow clusters and $d_f\approx1.73$ for (b) fast clusters.  The
    inset of (a) shows the fractal dimension $d_f$ for clusters of
    slow particles as a function of $\phi$.  }
  \label{clusters-VF805}
\end{figure} 

\medskip To characterize all of the clusters of slow/fast particles
that are present at a given time we show $R_g(N_c)$ for $\phi=0.805$
in Fig.~\ref{clusters-VF805}(a) for slow clusters and in
Fig.~\ref{clusters-VF805}(b) for fast clusters. We notice that there
are many slow clusters with $N_c > 10^3$, and even some with $N_c >
10^4$.

The relationship between the
radius of gyration and the cluster size can be described by a power
law of the form $R_g\propto{N_c}^{1/d_f}$, where $d_f$ corresponds to
the fractal dimension of the clusters, i.e., characterizes 
the shape of the clusters. For the packing fraction shown
in Fig.~\ref{clusters-VF805} we obtained a value $d_f\approx 1.91$ for
slow clusters and $d_f \approx 1.73$ for fast
clusters. 

The inset of Fig.~\ref{clusters-VF805}(a) shows the fractal dimension $d_f$
as a function of the packing fraction $\phi$, for slow 
clusters~\cite{footnote1}.
For slow clusters the value of $d_f$ increases with $\phi$
from a value $d_f \approx 1.65$ for $\phi = 0.60$ to a value $d_f
\approx 1.91$ for $\phi = 0.805$. 
This result suggests that the slow
clusters exhibit a more compact shape when approaching dynamical
arrest, as also suggested by Fig.~\ref{SD-2}(c). 
\begin{figure}[ht]
  \begin{center}
   \begin{picture}(220,320)
  \put(-4,150){\label{fig:dist-N-slow} \includegraphics[scale=.65]{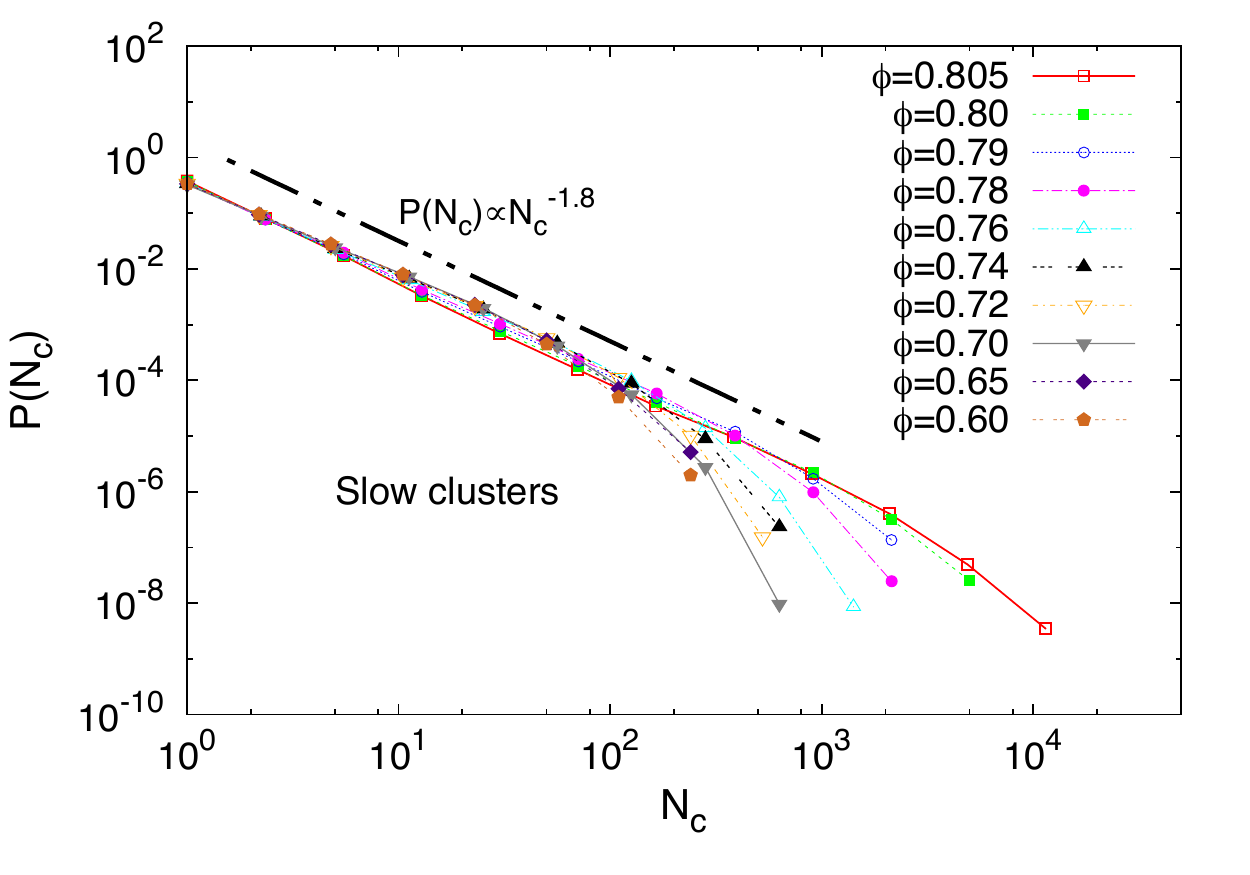}}
 \put(0,-10){\label{fig:dist-N-fast} \includegraphics[scale=.645]{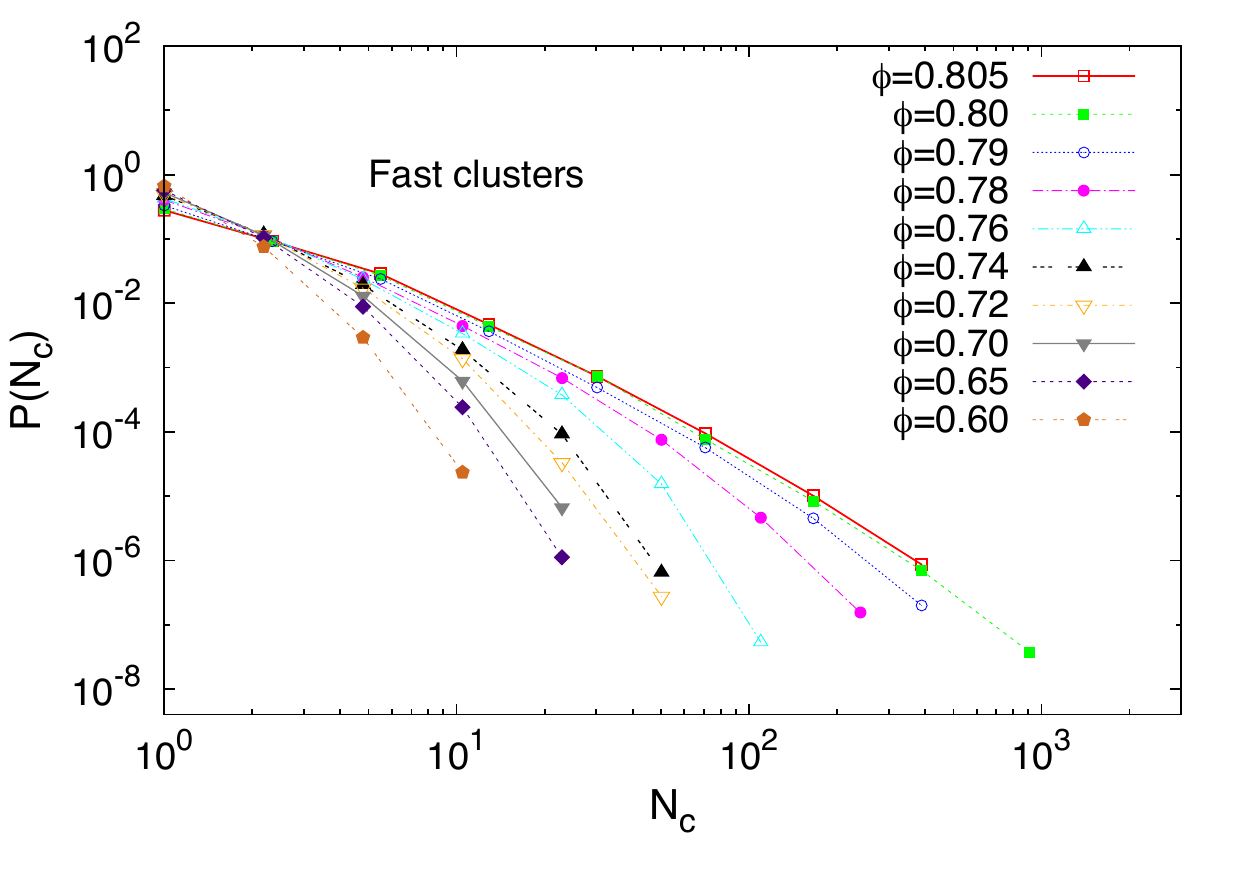}} 
 \put(40,295){\large{(a)}}
  \put(40,130){\large{(b)}}
   \end{picture}
 \end{center}
 \caption{Cluster size distribution $P(N_c)$ for
   different packing fractions $\phi$. (a) The distribution $P(N_c)$
   of slow clusters follows a power law $P(N_c)\propto {N_c}^{-1.8}$
   up to the cluster size $N^{*}$, which increases with increasing
   $\phi$. (b) Distribution of fast clusters.}
  \label{dist}
\end{figure} 
\begin{figure}[ht]
  \begin{center}
    \label{fig:rhoNc-collapse}  
    \includegraphics[scale=.655]{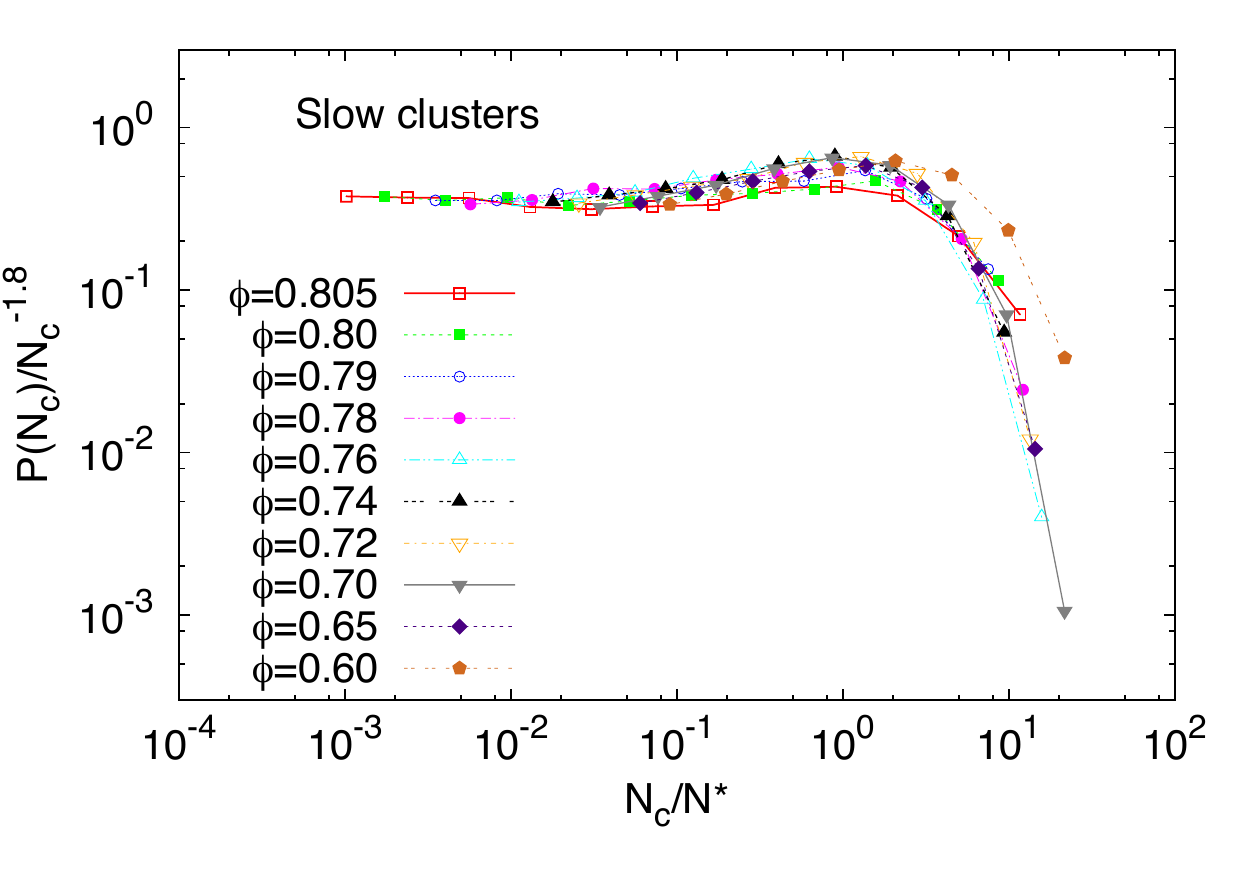}\\
    % \subfloat{\label{fig:rhoNc-collapse}  \includegraphics[scale=.655]{figures/dist-N-slow-collapse.pdf}}\\
 \end{center}
  \caption{Collapse of the data for slow clusters using the scaling ansatz $P(N_c)\sim N_c^{-\kappa}f(N_c/N^{*})$ with
   $N^{*}\propto (\phi_c-\phi)^{-1/\rho}$.
  } 
  \label{dist-collapse}
\end{figure} 

\begin{figure}[ht]
  \begin{center}
    \includegraphics[scale=.68]{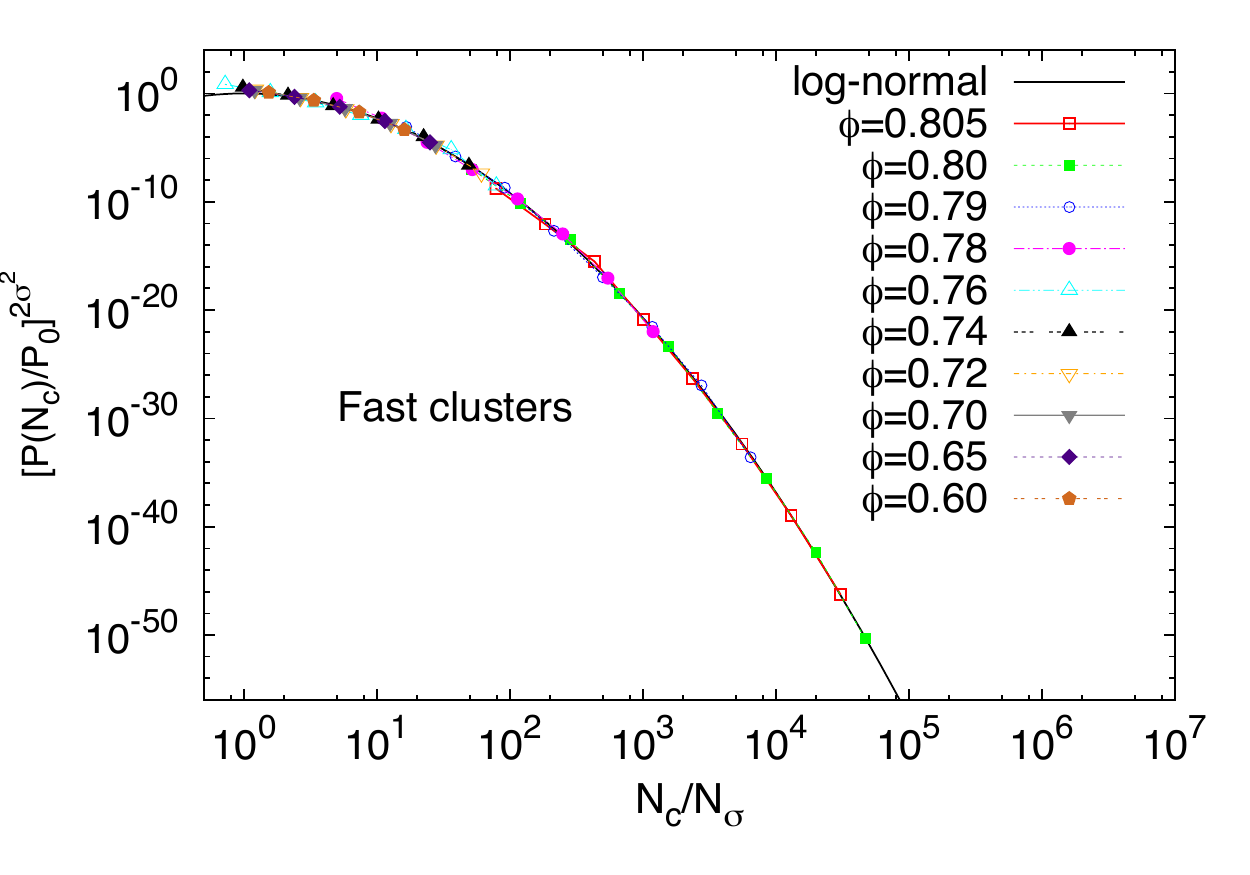}\\
    % \subfloat{\includegraphics[scale=.68]{figures/new-dist-N-fast_fit.pdf}}\\
 \end{center}
  \caption{Scaling plot for the fast cluster size
    distribution using Eq.~(\ref{PofNcfastfit-eq}). }
  \label{fig:lognormal-scaled}
\end{figure} 

\medskip We now turn to the distribution of cluster sizes $P(N_c)$
defined as the fraction of clusters of size $N_c$ (note that this is
not the standard definition in percolation theory). In
Fig.~\ref{dist}(a), $P(N_c)$ for slow clusters is shown for $0.60 \le
\phi \le 0.805$. As the packing fraction increases, $P(N_c)$
approaches a power law form $P(N_c) \propto N_c^{-\kappa}$ up to a
cutoff $N^{*}$ that grows with $\phi$. Fig.~\ref{dist-collapse} shows
$P(N_c)/N_c^{-\kappa}$ as a function of $N_c/N^{*}$, where $N^{*}
\propto (\phi_c-\phi)^{-1/\rho}$. Good collapse of the data is found
for all packing fractions except the lowest one, with the following
values for the fitting parameters: $\kappa \approx 1.8$ and $\rho \approx
1.63$. ($\phi_c = 0.82$ was previously determined from the simultaneous fit
of $\tau_{\alpha}$, $\chi_4$ and $\xi$). This suggests that the probability
distribution has the scaling form
\begin{equation}
P(N_c)\sim N_c^{-\kappa}f(N_c/N^{*})
\label{PofNcslowfit-eq}
\end{equation} 
and that $N^{*} \to \infty$ as $\phi \to \phi_c$ so that the
distribution becomes a simple power law $P(N_c)\propto {N_c}^{-\kappa}$.

By contrast, $P(N_c)$ for fast clusters, shown in Fig.~\ref{dist}(b),
does not exhibit a power law behavior: there is clearly curvature in
the log-log plot. However, a parabolic shape in the log-log plot
corresponds to a log-normal distribution 
\begin{equation}
P(N_c) \propto N_c^{-1}
    \exp\{-[\ln(N_c)-\ln(N_0)]^2/(2 \sigma^2)\}  \hspace*{10mm}\mbox{.}
\label{PofNcloglog-eq}
\end{equation} 
Rearranging Eq.~(\ref{PofNcloglog-eq}), we obtain
\begin{equation}
  \ln\{[P(N_c)/P_0]^{2 \sigma^2}\} = -
    [\ln(N_c/N_{\sigma})]^2 \hspace*{10mm}\mbox{,}
\label{PofNcfastfit-eq}
\end{equation}
where $P_0$ is a normalization constant, and $N_{\sigma} = N_0
\exp(-\sigma^2)$. We have successfully fitted $P(N_c)$ for each
packing fraction to Eq.~(\ref{PofNcfastfit-eq}) where $P_0$,
$N_{\sigma}$ and $\sigma$ are $\phi$-dependent fitting parameters.
Fig.~\ref{fig:lognormal-scaled} shows a scaling plot of
$[P(N_c)/P_0]^{2 \sigma^2}$ as a function of $N_c/N_{\sigma}$, where
very good collapse is obtained for the data for all packing fractions.
The log normal distribution for cluster sizes is the generic outcome
of coalescence growth mechanisms~\cite{Villarica1993}. It has been
reported in various studies of elemental clusters, both metallic and
non-metallic~\cite{Villarica1993, Buhrman1976, Wang1994, Mendham2001},
but also in such dissimilar cases as the distribution of sizes of
globular cluster systems in elliptical galaxies~\cite{Vesperini2000}. 

\subsection{Dynamic Susceptibility $\chi_4(t)$ and four-point structure function $S_4(q,t)$}\label{chi}
\subsubsection*{$\chi_4(t)$ as a function of packing fraction}
In the previous section we characterized spatial dynamic
heterogeneities by directly analyzing the size and shape of slow and
fast particles. A more common analysis of spatial dynamic
heterogeneities is indirectly via four-point correlation 
functions~\cite{Toninelli2005,Parsaeian2008,Flenner2011}. We
take this route in the following paragraph.

We begin with the dynamic susceptibility
\begin{equation}
\chi_4(t)=\overline{N[\left<Q^2_{\bf r}(t;t_0)\right>-\left<Q_{\bf
      r}(t;t_0)\right>^2]}.
\label{chi_4-eq}
\end{equation}
This quantity gives a global measurement of the fluctuations, and can
be interpreted as being proportional to the number of correlated slow
particles.
\begin{figure}
 {\includegraphics[width=0.47\textwidth]{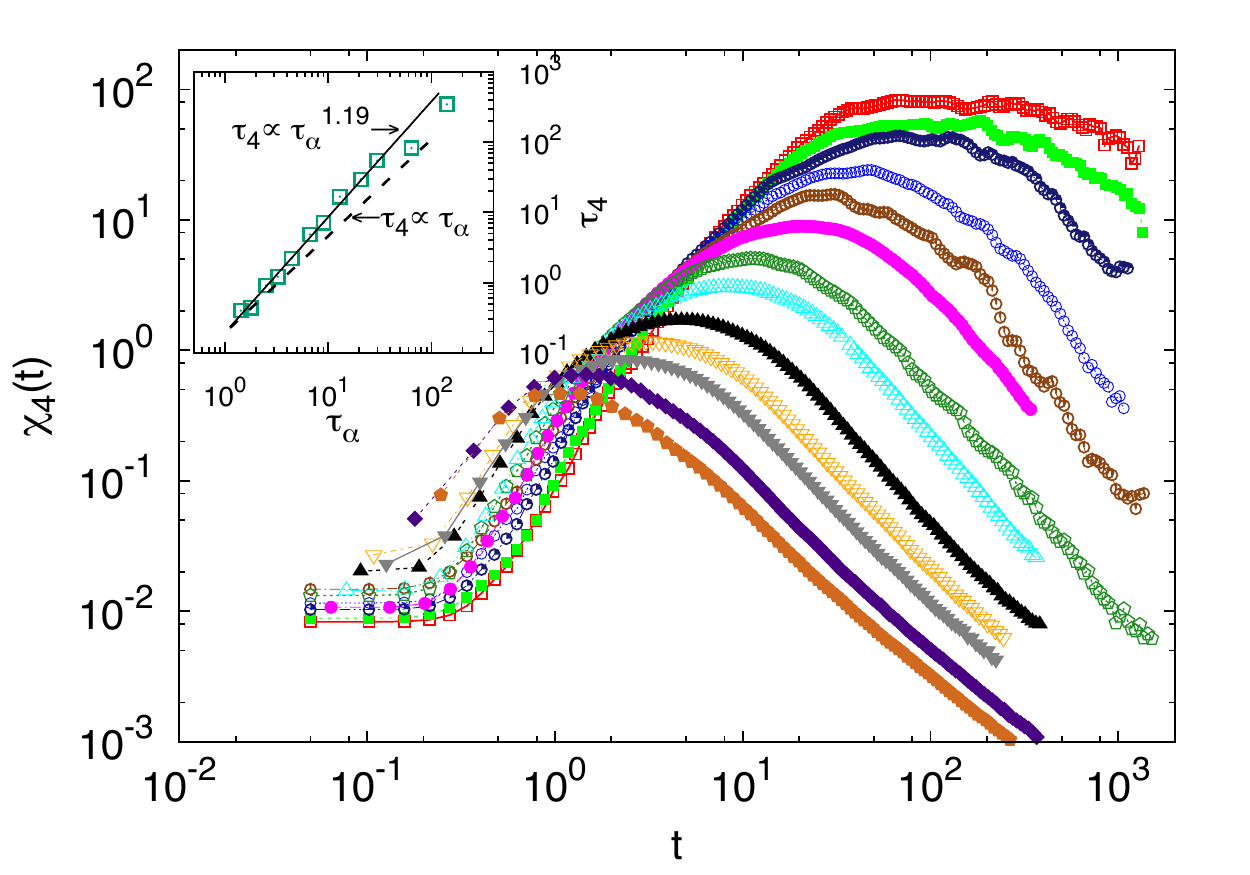}}
 \caption{Dynamic susceptibility as defined in
   Eq.~(\ref{chi_4-eq}) as function of time $t$ for various $\phi$
   (colors and symbols as in Fig.~\ref{Q-t}). Inset: $\tau_4$ against
   $\tau_{\alpha}$ for different packing fractions. The fit $\tau_4
   \propto \tau_{\alpha}^{1.19}$ excludes packing fractions
   $\phi>0.80$.}
     \label{chi4.pdf}
\end{figure} 
%----

Fig.~\ref{chi4.pdf} shows $\chi_4(t)$ for various packing fractions
$\phi$ with fixed $a=0.6$~\cite{footnote2}. The dominant features are a) a strong increase of
the peak value, $\chi_4^P$, as $\phi_c$ is approached, indicating a
strong increase in the number of correlated particles and b) a
correspondingly strong increase of the time, $\tau_4$, when the peak
occurs. The latter is in agreement with the slowing down of the
dynamics, discussed in Sec.~\ref{sec:overlap} for the dynamic
overlap. In fact the time $\tau_4$ is related to the relaxation time
$\tau_{\alpha}$ via a power law (see inset of Fig.~\ref{chi4.pdf}).
For the former, we had shown in~\cite{AvilaPRL2014} that
$\chi_4(\tau_{\alpha}) \propto \left ( \phi_c - \phi \right
)^{-\gamma_{\chi}}$ with $\phi_c \approx 0.82$ and $ \gamma_{\chi}
\approx 2.5$.

\subsubsection*{Four-Point Structure Factor}\label{S4xi-sec}

The spatially resolved fluctuations of the overlap can be studied with the help of the four--point structure factor $S_4(q,t)$ given by 
\begin{eqnarray}\label{s4-eq}
\lefteqn{S_4(q,t)/N =} \\[0.5ex]
 & &  \left \{ \overline{ 
    \left[\left\langle W_{\bf r}({\bf q},t;t_0) 
      W_{\bf r}({\bf -q},t;t_0) \right\rangle 
      - \left\langle W_{\bf r}({\bf q},t;t_0) 
      \right\rangle \left\langle 
      W_{\bf r}(-{\bf q},t;t_0)
      \right\rangle  \right]}\right \} , \nonumber
\end{eqnarray}
where
\begin{equation}
W_{\bf r}({\bf q},t)=\frac{1}{N}\sum_{i=1}^{N_{\bf r}}
\exp{[i{\bf q}\cdot {\bf r}_i(t_0)] \theta(a-|{\bf r}_i(t_0+t)-{\bf
    r}_i(t_0)|)}. 
 \end{equation}
Here $\{{\cdots }\}$ denotes an average over wave vectors $\bf{q}$ of
fixed magnitude $|{\bf q}|=q$.

\begin{figure}[h]
  { \includegraphics[width=.46\textwidth]{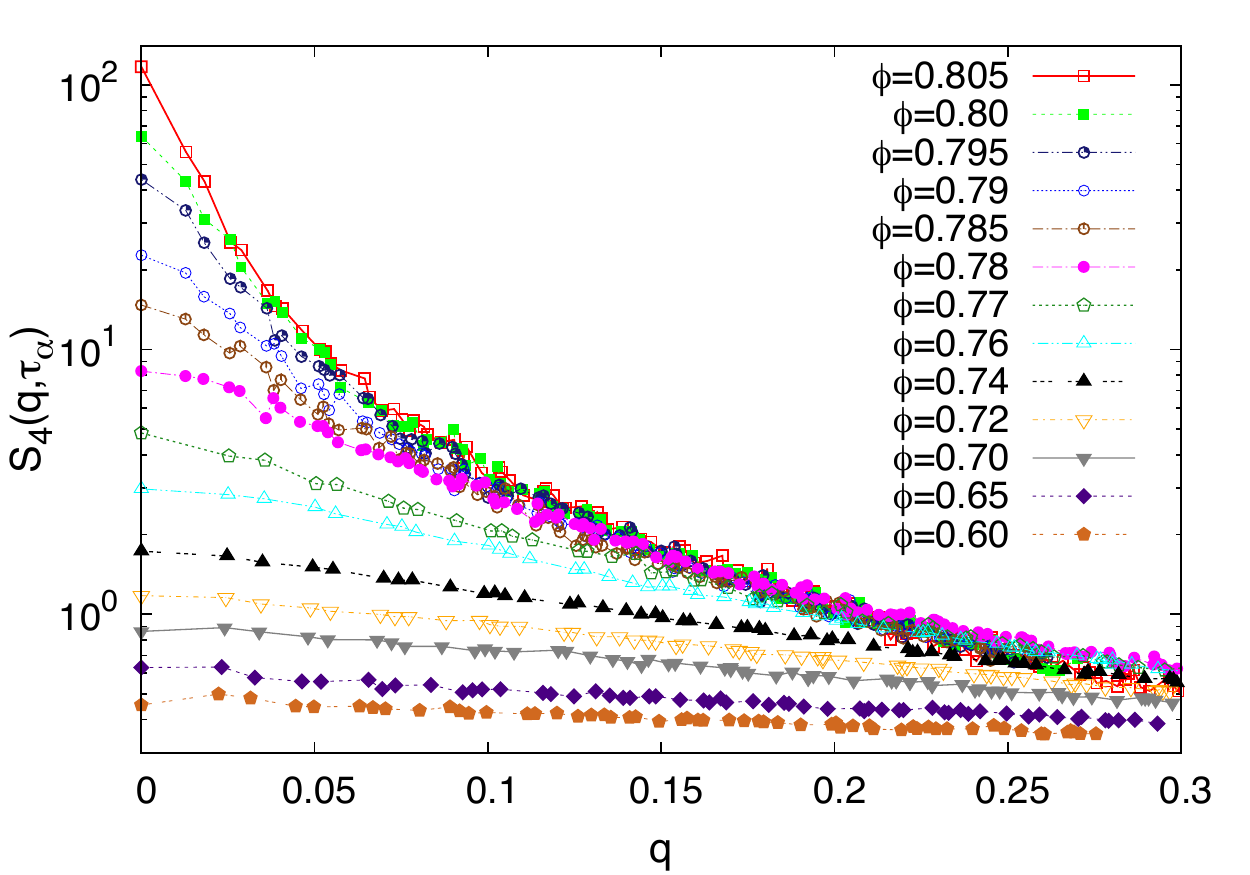}}
  \caption{Four-point structure factor
    $S_4(q,\tau_{\alpha})$ for different packing fractions. The
    values at $q=0$ were obtained by the direct calculation of
    Eq.~(\ref{s4-eq}). }
  \label{S4}
\end{figure} 

\begin{figure}[h]
\includegraphics[scale=.65]{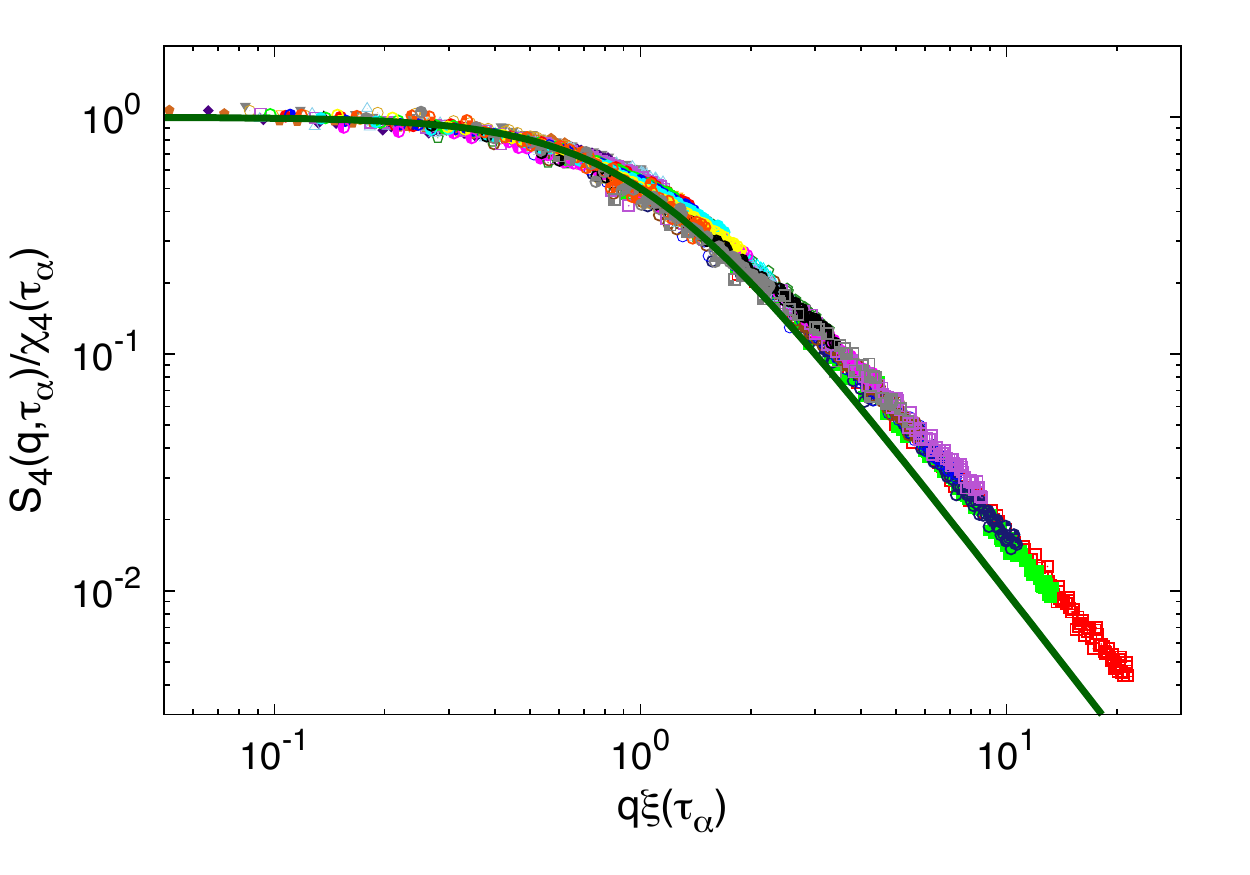}
\caption{Scaling plot of $S_4(q,\tau_{\alpha})$ for all simulated
  values of $\phi$ and $\varepsilon$ (see Sec.~\ref{epsilon_varied}
  below). }\label{scalingS4}
\end{figure}

The four-point structure factor $S_4(q,\tau_{\alpha})$, evaluated at
the $\alpha$-relaxation time, is displayed in Fig.~\ref{S4} for
various $\phi$. We observe a strong increase of $S_4(q,\tau_{\alpha})$
for small wavenumber as $\phi \to \phi_c$, which is to be expected
since 
\begin{equation}\label{qto0limit}
\lim_{q\rightarrow0}S_4(q,t)=\chi_4(t). 
\end{equation}
The data for all investigated packing fractions can be collapsed to a
single curve, when plotting $S_4(q,\phi)/\chi_4(\phi)$ as a function of
$q \xi(\phi)$, with all quantities evaluated at $t=\tau_{\alpha}$ (see Fig.~\ref{scalingS4}). For small wavenumber
the scaling function is well approximated by an Ornstein-Zernike (OZ)
fit,
\begin{equation}
S_4(q,\tau_{\alpha})=\frac{\chi_4(\tau_{\alpha})}{1+[{q}\xi(\tau_{\alpha})]^2},
\label{S4-OZ}
\end{equation}
which allows us to extract the correlation length as a function of $\phi$.
We had shown in~\cite{AvilaPRL2014} that the resulting 
correlation length, $\xi(\tau_{\alpha})$, 
as function of $\phi$ diverges as a power law,
\begin{equation}
\xi(\tau_{\alpha}) \propto \left ( \phi_c - \phi \right )^{-\gamma_{\xi}},
\end{equation}
with $\gamma_{\xi}=1.6$. %
Using a power law fit for $\tau_{\alpha}$ then implies an algebraic growth of the correlation length with relaxation time: $\tau_{\alpha} \propto \left [
  \xi(\tau_{\alpha}) \right ]^z$ where $z=\gamma_{\tau}/\gamma_{\xi}$.
Such an algebraic dependence is in contrast to most 3D non--dissipative glasses
where instead of an algebraic an exponential dependence prevails, but in agreement with a recent study~\cite{FlennerSzamel2015} of 2D glasses.

\begin{figure}
  \begin{center}
  \includegraphics[scale=.645]{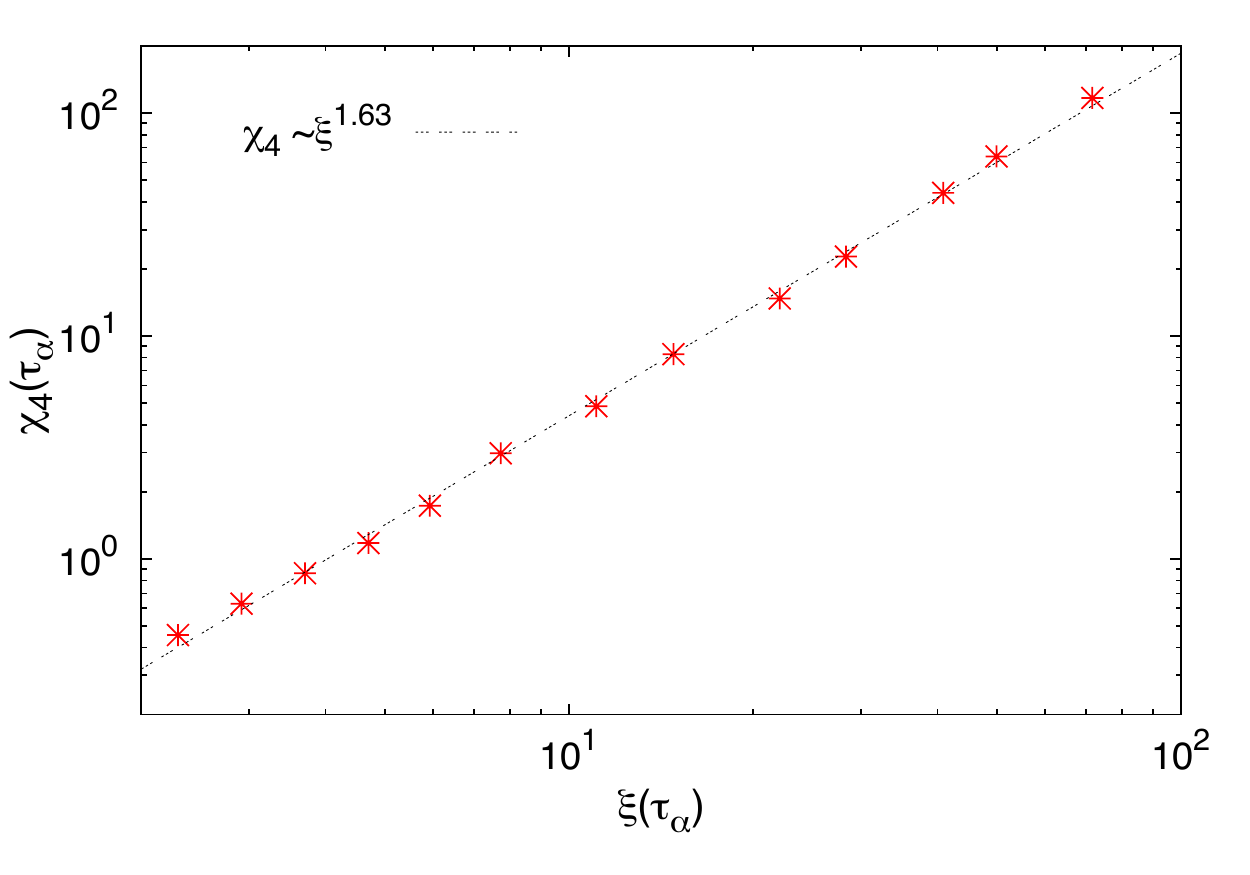}
  \end{center}
  \caption{$\chi_4(\tau_{\alpha})$ against
    $\xi(\tau_{\alpha})$. The dashed line corresponds to the fit
    $\chi_4(\tau_{\alpha})\propto \xi^{d_f'}(\tau_{\alpha})$, with
    ${d_f'}\approx1.6$.}  
    \label{xi-chi_4-2}
\end{figure}

As already mentioned, $\chi_4(t)$ measures the number of particles
moving together in a cooperative manner, and $\xi(t)$ is a measure for
the spatial extension of these cooperative regions. Thus, from the
relationship $\chi_4(t)\propto \xi^{d_f'}(t)$ the exponent $d_f'$ is
usually interpreted as the fractal dimension of the
clusters~\cite{Berthier2011}. For instance, this would mean that the
case of $d_f'=d$ corresponds to compact clusters, whereas the case of
$d_f'=1$ corresponds to strings. The scaling behavior of
$\chi_4(\tau_{\alpha})$ versus $\xi(\tau_{\alpha})$ for this system is
shown in Fig.~\ref{xi-chi_4-2}, with the fitted value $d_f'\approx1.6$.
This is close to the value determined from the radius of gyration for
$\phi\sim 0.6$, but fails to show the density dependence which we
detected in Sec.~\ref{cluster-sec}. There we looked at clusters of a
specific size, $N_c$, and determined their radius of gyration. Here
the relation is less clear, because $\chi_4$ corresponds to an
average or typical cluster size.

An alternative explanation for the observed scaling $\chi_4(t)\propto
\xi^{d_f'}(t)$ has been suggested, namely that correlated regions could
be compact, but their sizes could have a wide distribution.  The OZ
form of $S_4(q,t)$ implies a sufficiently fast decay of $G_4(r,t)$ for
large distances $r$ and hence is not compatible with a wide
distribution on the {\it largest scales}. However, we have seen in
Sec.~\ref{cluster-sec} that the cluster size distribution indeed becomes
increasingly wider as $\phi\to \phi_c$: The distribution decays
algebraically, crossing over to an exponential at the cutoff $N^*$
which diverges as $\phi\to \phi_c$. Hence it becomes increasingly
difficult to disentangle the two effects, namely the clusters
compactifying and the distribution widening. Since the effects work in
opposite directions, they might partially compensate. In any case the
estimates of the fractal dimension directly from clusters of a specific
size is superior to the rather indirect way using the relation between
$\chi_4$ and $\xi$. The latter invariably gives rise to a constant value of
$d_f'$ as long as both quantities follow power laws with density
independent exponents.

It is important to remark that, as demonstrated in
Ref.~\cite{Lebowitz,Berthier2005,Flenner2011}, the equivalence of $\chi_4(t)$ and the limit
$S_4(q \to 0,t)$ is a subtle point. For numerical simulations where the
particle density and the relative concentrations of each particle type
are held fixed, additional contributions are required to relate the two
quantities~\cite{Berthier2005, Flenner2011}.
In this work, since we cut the simulated box into sub--boxes of equal
size, the number of particles and particle concentration varies with time and
between sub--boxes. Therefore, all fluctuations are expected to be
accounted for in Eq.~(\ref{chi_4-eq}),
and no additional terms are needed
in the calculation of $\chi_4(t)$. This means that the $q\rightarrow
0$ limit of $S_4(q,t)$ is expected to be well described by $\chi_4(t)$
obtained from the calculation of Eq.~(\ref{chi_4-eq}), whereas for other
ensembles, $S_4(q=0,t)$ is obtained by an extrapolation of
$\lim_{q\rightarrow0} S_4(q,t)$~\cite{Glotzer2003,Flenner2011} or by
directly calculating the missing contribution to $\chi_4(t)$ as
presented in Ref.~\cite{Flenner2011} (see Appendix). %

\subsubsection*{Finite size analysis of $\chi_4$}
 It is clear from Eq.~(\ref{chi_4-eq}) that the
dynamic susceptibility is a product of the variance of the overlap
$Q_{\bf r}(t;t_0)$, with a factor of $N$, which makes it scale like a
constant as a function of $N$ in the thermodynamic limit $N \to
\infty$. 

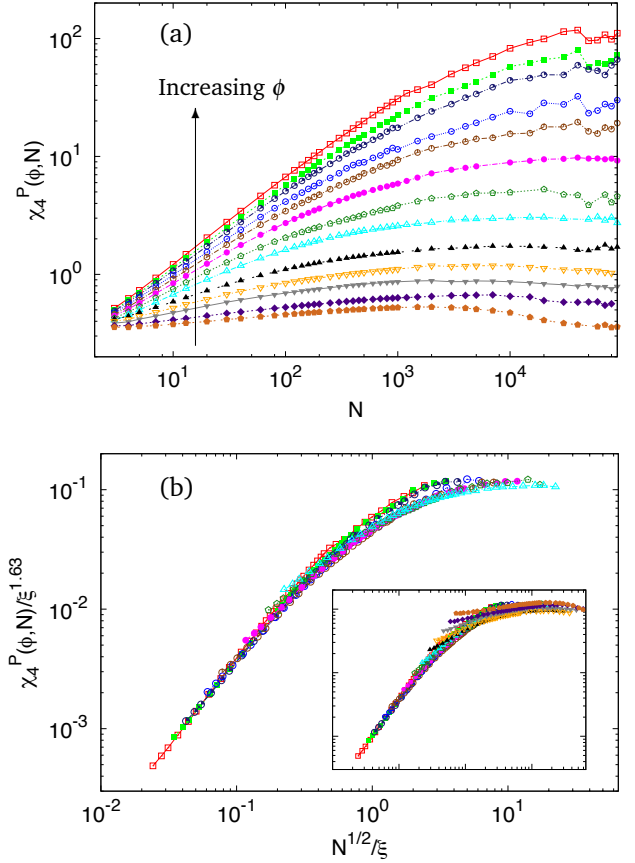
\begin{figure}
  \begin{center}
    \begin{picture}(220,320)
  \put(-2,155){\includegraphics[scale=.67]{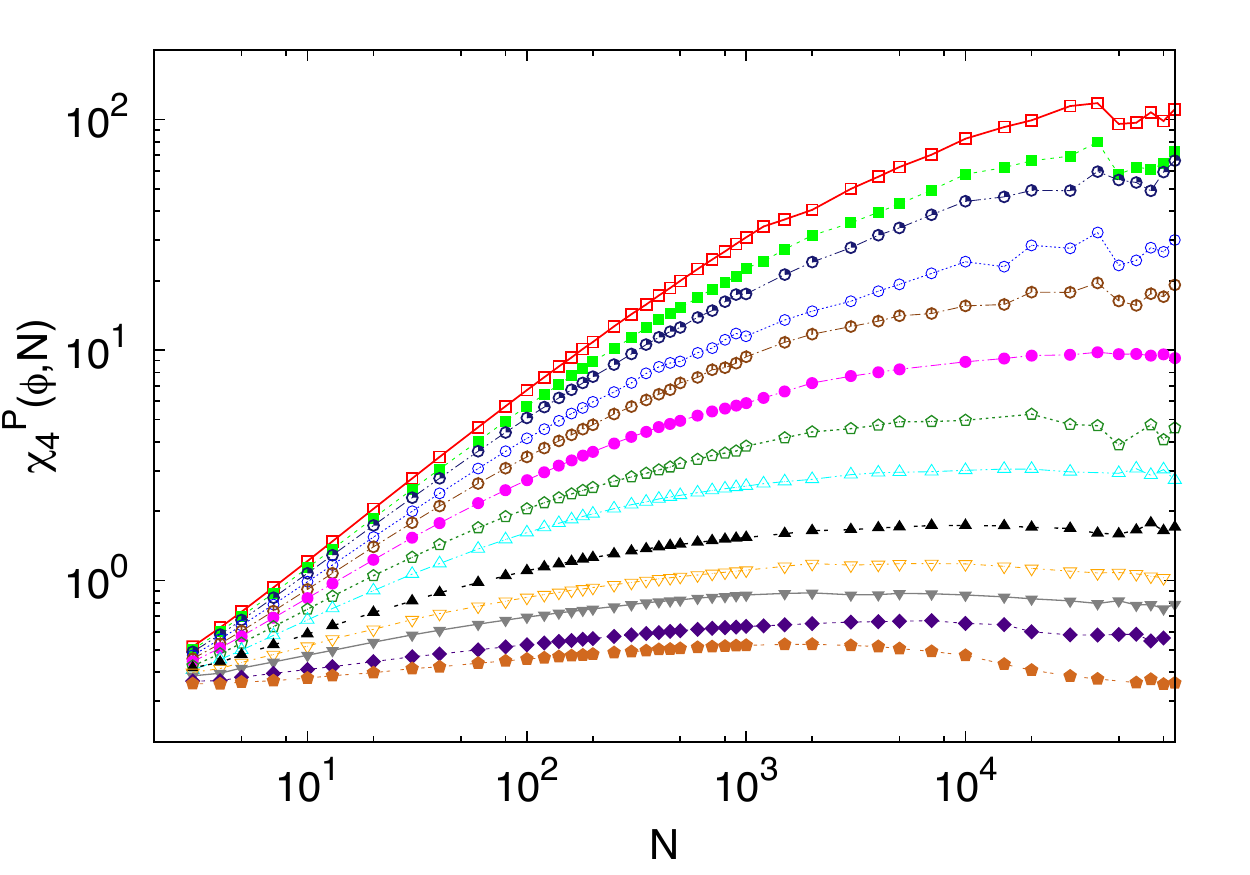}}
 \put(-5,-18){\includegraphics[scale=.68]{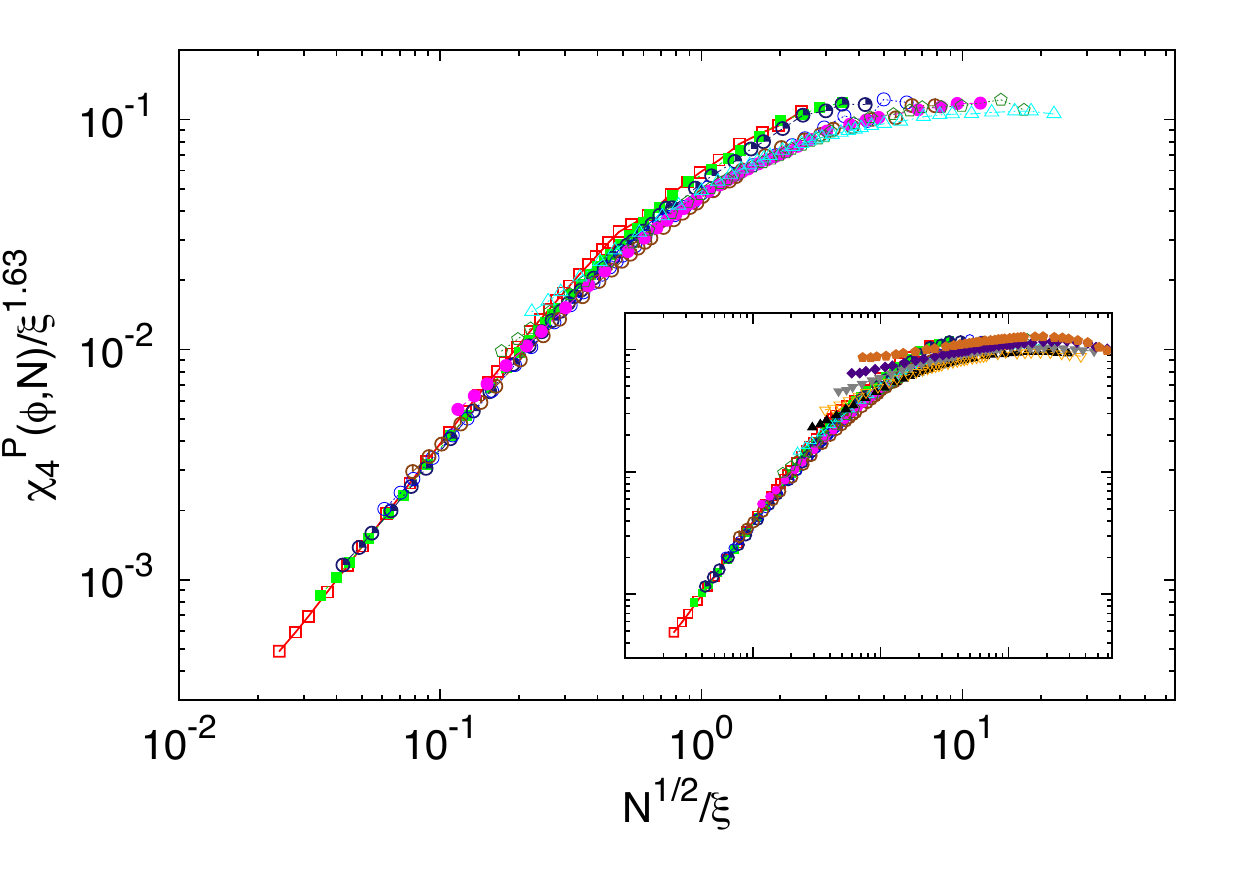}} 
 \put(52,300){\large{(a)}}
  \put(52,280){{Increasing $\phi$}}
   \put(66,185){\vector(0,1){90}}
  \put(52,128){\large{(b)}}
 \end{picture}
  \end{center}
  \caption{(a) Maximum value $\chi_4^P$  of the 
            dynamic susceptibility as a function of the number of particles $N$
            in the analyzed subbox. (b) Finite size scaling plot 
             $\chi_4^P \xi^{\gamma_{\chi}/\gamma_{\xi}}(\sqrt{N}/ \xi)$
             with $\xi$, $\gamma_{\chi}$, and $\gamma_{\xi}$ as described in the
             text for packing fractions $\phi\geq 0.76$. Inset: Same quantities and axis range as in main panel for all packing fractions.
    } 
  \label{chi4P}
\end{figure}

The maximal value $\chi_4^P$ corresponds to a maximal number of
correlated slow particles and is dependent on $\phi$ and the number of
particles in the analyzed sub--box (see Sec.~\ref{avg}) as shown in
Fig.~\ref{chi4P}(a). $\chi_4^P$ is observed to increase with system
size as long as the system size is smaller than the correlation length
and saturates once the system size is comparable or larger than the
correlation length. These observations can be quantified using finite
size scaling. The data for different $\phi$ can be collapsed
approximately to a single function $\chi_4^P
\xi^{\gamma_{\chi}/\gamma_{\xi}}(\sqrt{N}/ \xi)$ as shown in
Fig.~\ref{chi4P}(b)~\cite{footnote3}, where $\xi$, $\gamma_{\chi}$, and $\gamma_{\xi}$
were determined from $\chi_4$ and $S_4(q,t)$, as discussed above.
Collapse of the data works well, in particular packing fractions $\phi
\geq 0.76$ (see main panel of Fig.~\ref{chi4P}(b)), collapse in the full range $0.01<\sqrt{N}/\xi <100$.  Data for
lower packing fractions deviate from the scaling function, when the
system size becomes comparable or larger than the correlation length
 (see inset of Fig.~\ref{chi4P}(b)).
To conclude, finite size scaling of $\chi_4^P$ is fully consistent
with the critical behavior extracted from $\chi_4(\tau_{\alpha})$ and
$S_4(q,\tau_{\alpha})$.

\subsubsection*{Cutoff Dependence}\label{a-dependence}

All previous results were calculated using $a=0.6$ (measured in units of $r_1$) in 
Eq.~(\ref{Q(t)-eq}). This value has been chosen by most studies~\cite{Glotzer2003,Flenner2011,Karmakar2009} 
which aim to calculate the extent of the
dynamical heterogeneities and in particular $\xi(t)$. However, the dependence
of the correlation length $\xi(t)$ on the parameter $a$ has not been
explored in detail.
 
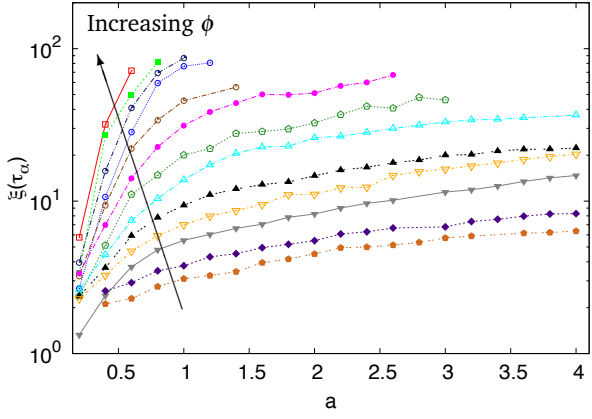
\begin{figure}[ht]
  \begin{center}
   \begin{picture}(160,160)
 \put(-33,-10){\includegraphics[scale=.65]{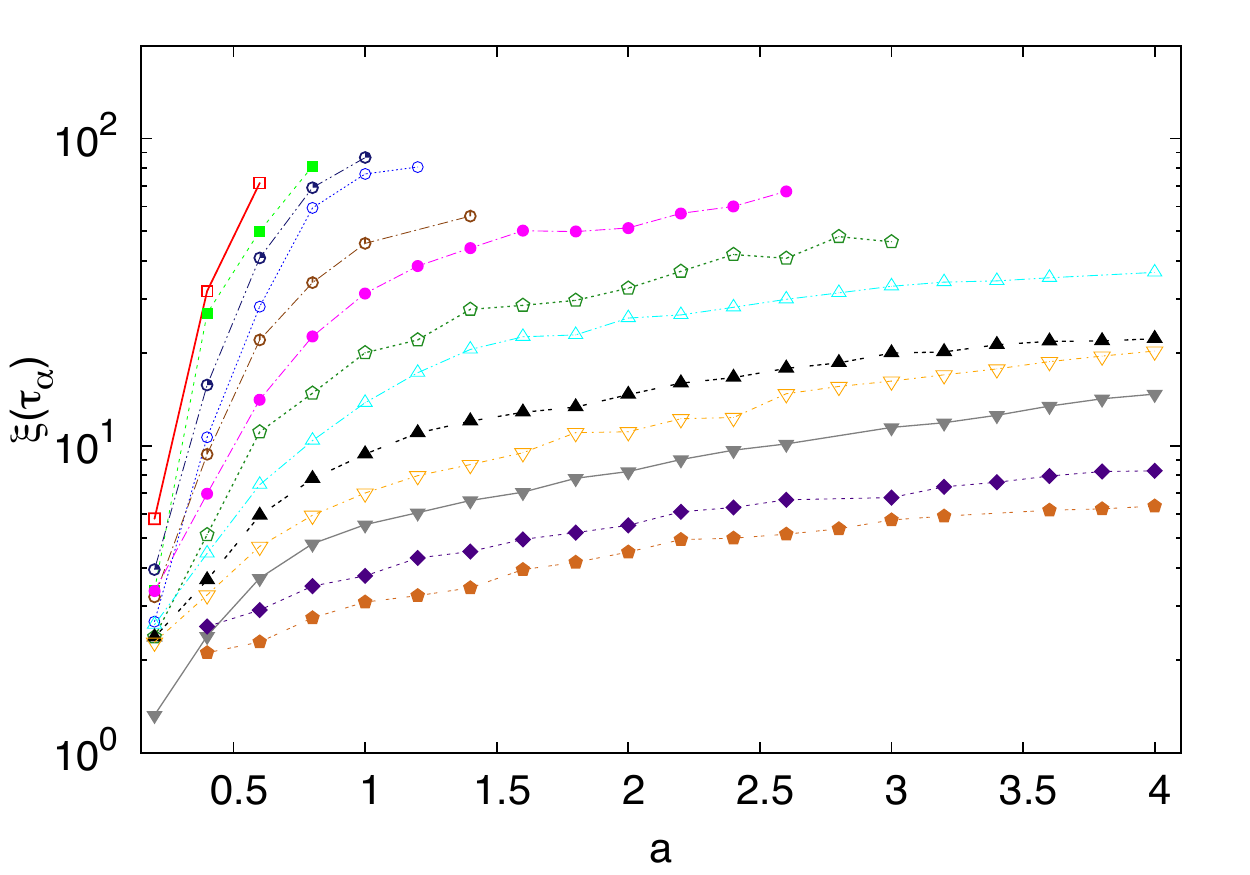}}
  \put(-1,135){{Increasing $\phi$}}
   \put(35,30){\vector(-1,3){32}}
  \end{picture}
  \end{center}
  \caption{$\xi({\tau_{\alpha}})$ against $a$ for
    $\phi=0.60$, $0.0.65$, $0.70$, $0.72$, $0.74$, $0.76$, $0.77$,
    $0.78$, $0.785$, $0.79$, $0.795$, $0.80$ and $0.805$ (from bottom
    to top). First, the grow of $\xi(\tau_{\alpha})$ occurs very
    rapidly for small $a$, then it goes through a crossover to finally
    go to a slower growth. } 
  \label{xi-vs-a}
\end{figure}

The growing behavior of $\xi(\tau_{\alpha})$ as a function of $a$ is
shown in Fig.~\ref{xi-vs-a} for the range $0.2\le a\le 4.0$. 
It can be observed that $\xi(\tau_{\alpha})$ goes through different
regimes. First a rapid increase can be identified for $a\lesssim
1$, then, a crossover, and finally, a much slower growth for
$a\gtrsim1.2$.

\begin{figure}[h!]
  \begin{center}
\begin{picture}(220,320)
  \put(-4,150){\includegraphics[scale=.645]{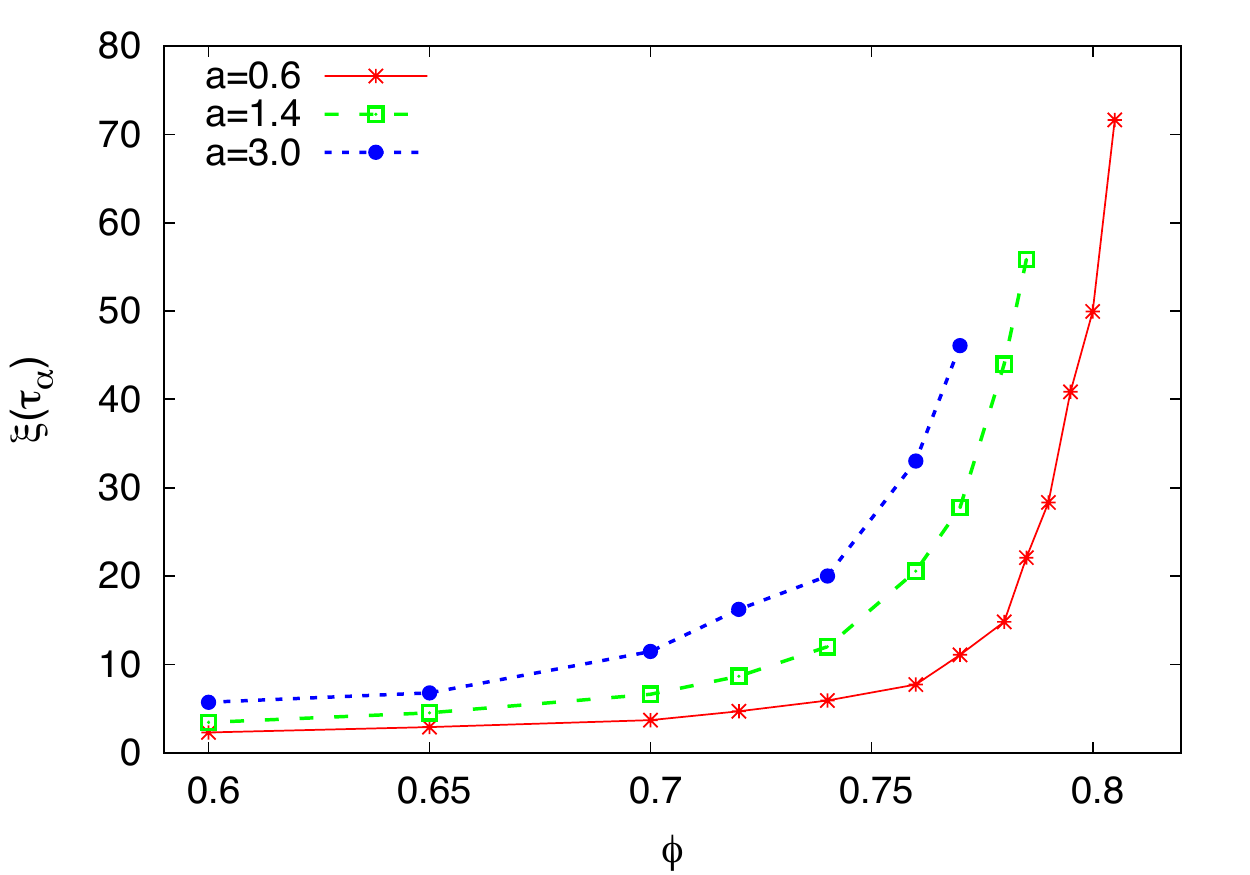}}
 \put(0,-10){\includegraphics[scale=.636]{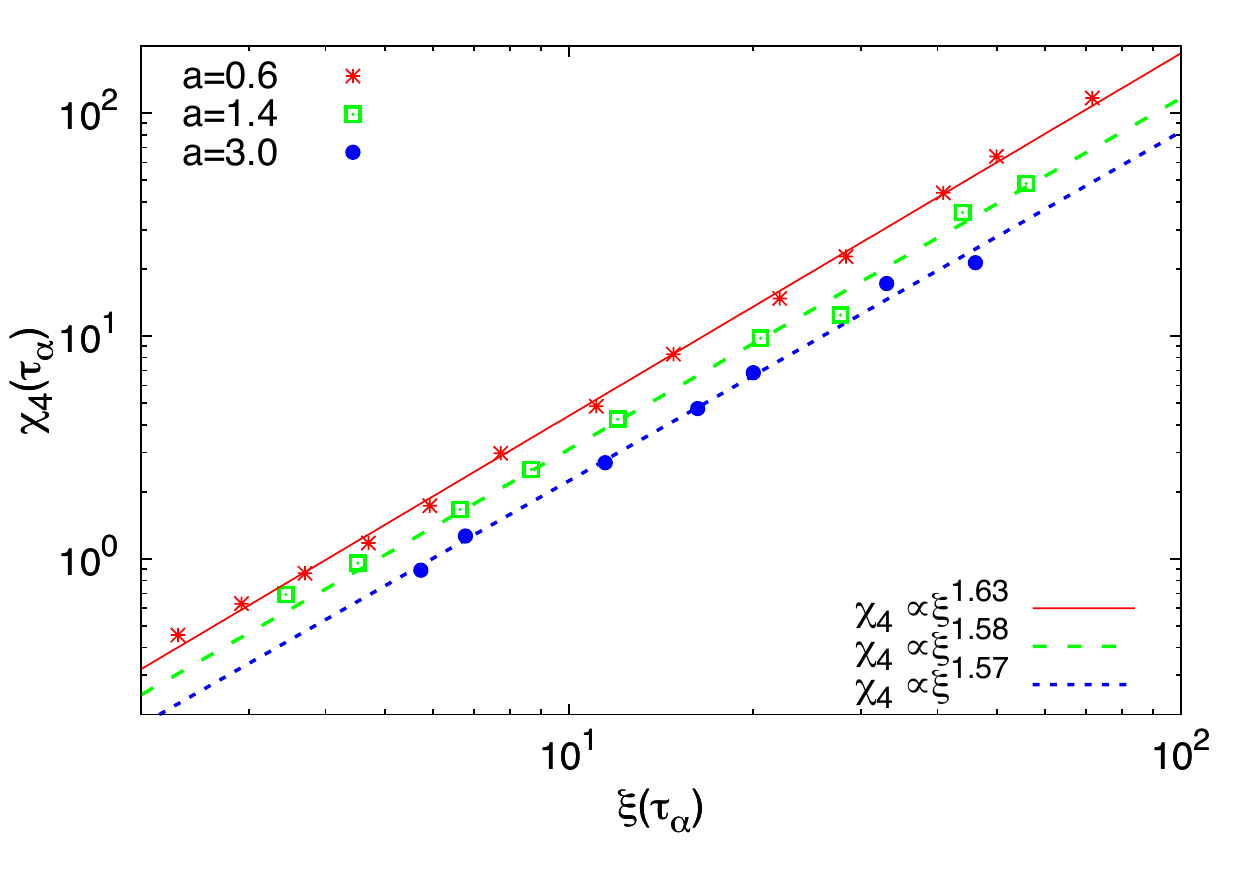}} 
 \put(50,210){\large{(a)}}
  \put(50,90){\large{(b)}}
 \end{picture} 	
   \end{center}
  \caption{(a) The dynamical correlation length
    $\xi(\tau_{\alpha})$ as a function of packing fraction $\phi$ for
    different values of the parameter $a$ of the overlap function. 
    (b) $\chi_4(\tau_{\alpha})$ against
    $\xi(\tau_{\alpha})$ for three different choices of the parameter
    $a$. The different lines correspond to the fit
    $\chi_4(\tau_{\alpha})\propto \xi^{d_f'}(\tau_{\alpha})$ for each
    $a$. We find similar fitted values of $d_f'$ for the three
    different values of $a$.} 
  \label{diff-a}
\end{figure}

In Fig.~\ref{diff-a}(a), we compare the correlation length as a function of $\phi$ for our
standard choice ($a = 0.6$) with two other choices of $a$, namely $a=1.4$ and
$a = 3.0$. All values for $\xi$ are obtained from the structure
function $S_4(q,\tau_{\alpha})$ with help of a fit to the OZ form.
In Fig.~\ref{diff-a}(a) a similar trend in the growing behavior of
$\xi(\tau_{\alpha})$ with $\phi$ can be seen for the different values
of $a$. Whereas for low packing fraction the points of $\xi(\tau_{\alpha})$
are very close to each other, the curves start to deviate
considerably from each other for high packing fractions. This suggests
that for the case of high packing fractions, when the heterogeneities
in the dynamics become more pronounced, the selection of $a$ has a bigger
impact on the result of $\xi$ than for low packing fractions, when the
dynamics is governed by collisions between pairs of particles.

With this in mind, it should be interesting to determine how the
relationship $\chi_4(\tau_{\alpha})\propto \xi^{d_f'}(\tau_{\alpha})$
changes with different values of $a$. It turns out that despite the
large difference in the values of $\xi$ for high packing fractions
with the choice of $a$, the relationship $\chi_4(\tau_{\alpha})\propto
\xi^{d_f'}(\tau_{\alpha})$ only changes by a multiplicative constant,
with the exponent $d_f'$ similar for all values of $a$ mentioned
before. These results are shown in Fig.~\ref{diff-a}(b) for the three
different values of $a$. These results indicate that independently of
the choice of $a$, $\chi_4(\tau_{\alpha})$ grows with
$\xi(\tau_{\alpha})$ in the same way. 

We also checked the fractal dimension obtained from the radius of 
gyration and again found it to be largely independent of $a$ in the 
range $0.6\leq a \leq 3$, for example for $\phi=0.76$, $1.78 \lesssim d_f \lesssim 1.84$.
This is slightly higher than the value found for $d_f'$.

\begin{figure}[h]
{\includegraphics[scale=.665]{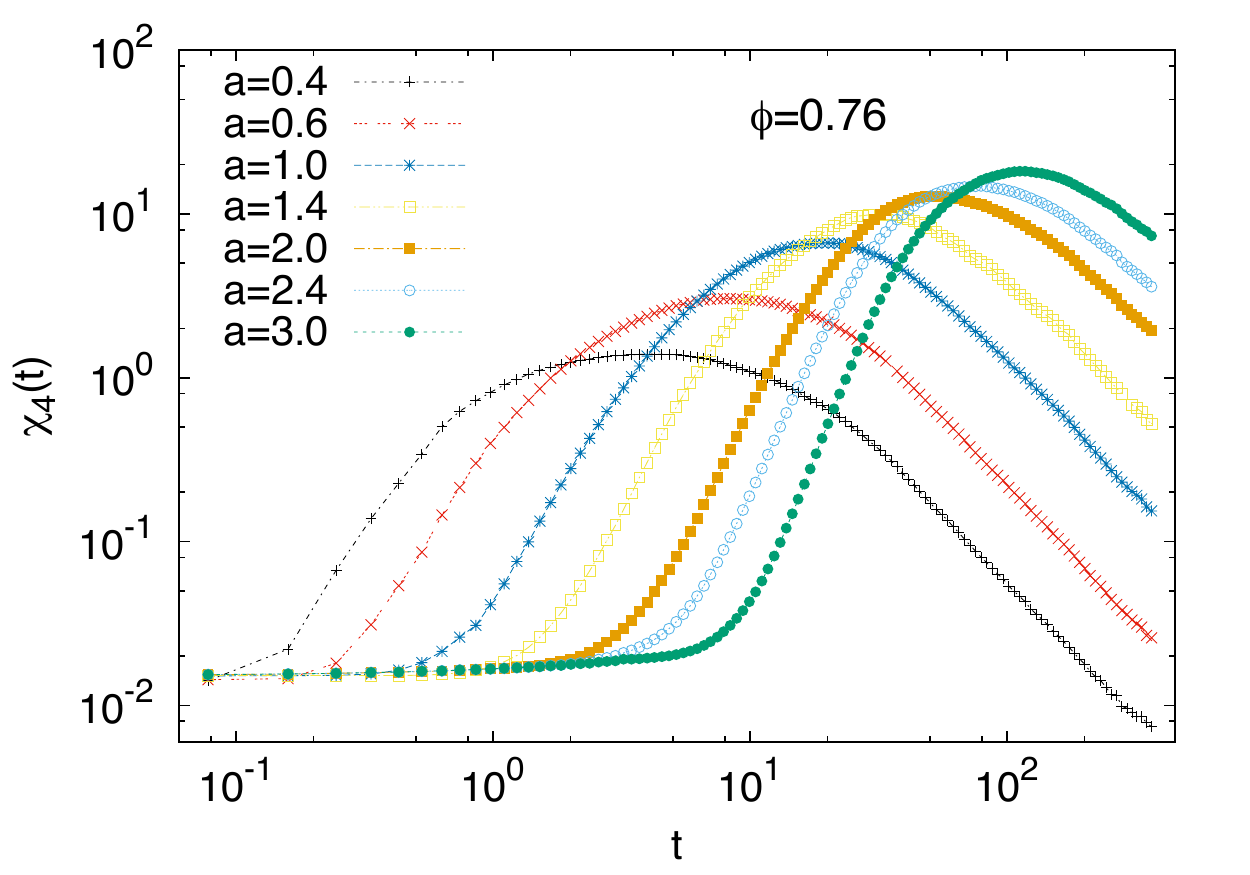}}
\caption{\label{chi4-diff-a}$\chi_4(\tau_{\alpha})$ for different values of the cutoff $a$.}
\end{figure}

It is known that the height of the peak of $\chi_4(t)$ increases with
increasing values of $a$ up to a certain value, which will be called
$a_{\text{max}}$. Then, for $a>a_{\text {max}}$ the peak is seen to
decrease~\cite{Abate2007,Glotzer2003}. We do observe the expected
increase of $\chi_4(t)$ for small $a$ (see Fig.~\ref{chi4-diff-a}) and
the data indicate saturation around $a=3$. 
Also, we see the shift of $\chi_4(t)$ 
to longer times with increasing $a$. However, for the packing
fractions studied in this work the value of $a_{\text{max}}$ was not
reached. In order to reach this value, the simulations would have to
be extended to much longer times. This is a difficult task considering
the large number of collisions involved. 
We therefore leave this for future work.

\subsubsection*{Dependence on inelasticity $\varepsilon$}
\label{epsilon_varied}
\begin{figure}
  \includegraphics[scale=.66]{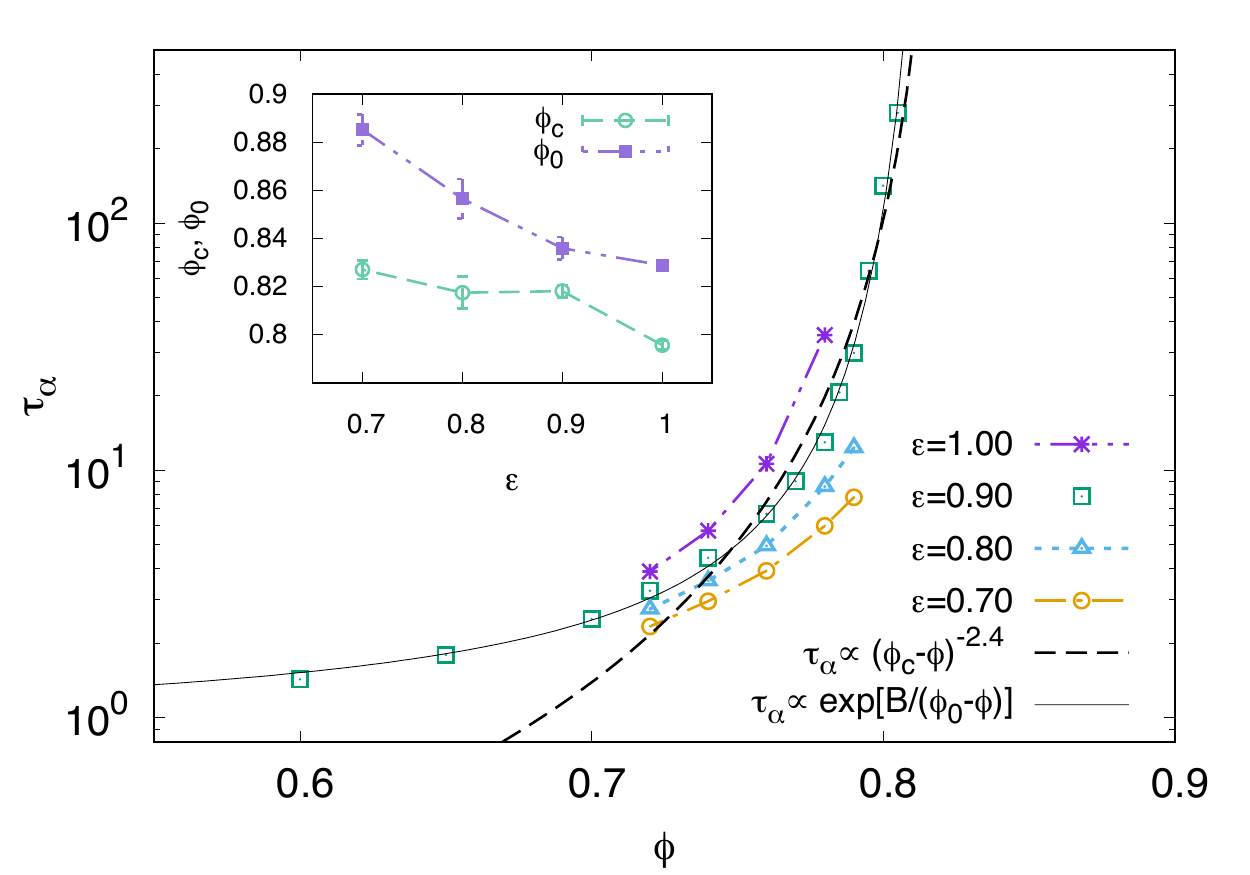}
  \caption{\label{tau}$\tau_{\alpha}$ as a function of $\phi$ for
    several values of $\varepsilon$; data fitted to $\tau_{\alpha}\propto (\phi_c-\phi)^{-\gamma_{\tau}}$ 
    (dashed line for $\varepsilon=0.90$) and $\tau_{\alpha}\propto \exp{[B/(\phi_0-\phi)]}$ 
    (solid line for $\varepsilon = 0.90$);
    the resulting values of $\phi_c(\varepsilon)$ and $\phi_0(\varepsilon)$ are shown in the inset.}
\end{figure}

Different granular materials are characterized by different
coefficients of restitution, determined by the microscopic properties
of the constitutive grains. $\varepsilon$ is the control parameter which determines
how non--dissipative a system is. Hence a question naturally arises: How
universal are our results with respect to variations in $\varepsilon$?
To answer this question, we present a few selected results:

The $\alpha$-relaxation time grows more slowly with packing fraction 
for the more inelastic system, indicating that $\phi_c$ increases with
decreasing $\varepsilon$ (see Fig.~\ref{tau}), as predicted by mode-coupling theory~\cite{Kranz2010}.
From the point of view of the fits discussed in Sec.~\ref{sec:overlap}, this corresponds to the parameters $\phi_c(\varepsilon)$ and $\phi_0(\varepsilon)$ being monotonous decreasing functions of $\varepsilon$, as shown in the inset of Fig.~\ref{tau}.

Similarly, the growing behavior of $\chi_4$ and $\xi$ is compatible with an $\varepsilon$--dependent 
critical density $\phi_c(\varepsilon)$ found in~\cite{Kranz2010} (not shown here). 
However, a robust
scaling law relates the two quantities, $\chi_4(\tau_{\alpha})\propto
\xi^{d_f'}(\tau_{\alpha})$, with an exponent $d_f' \sim 1.6$ which is
independent of $\varepsilon$ (see Fig.~6 in~\cite{AvilaPRL2014}).

The time-dependent overlap $Q(t)$ (Eq.~(\ref{Q_avg(t)})) for the range
$0.7\leq\varepsilon\leq 1.0$ can equally well be fitted to the
empirical form $Q^{-1}(t)=(t/\tau_0)^{\beta}+1$. In fact the data in
Fig.~\ref{Qofbetalntovertau0} include data sets for the above range of
$\varepsilon$. Similarly, data for $S_4(q,t)$ for different values of
$\varepsilon$ can be collapsed to the universal function shown in
Fig.~\ref{scalingS4}, in fact the data are included.

\begin{figure}[b!]
  \begin{center}
      \includegraphics[scale=.65]{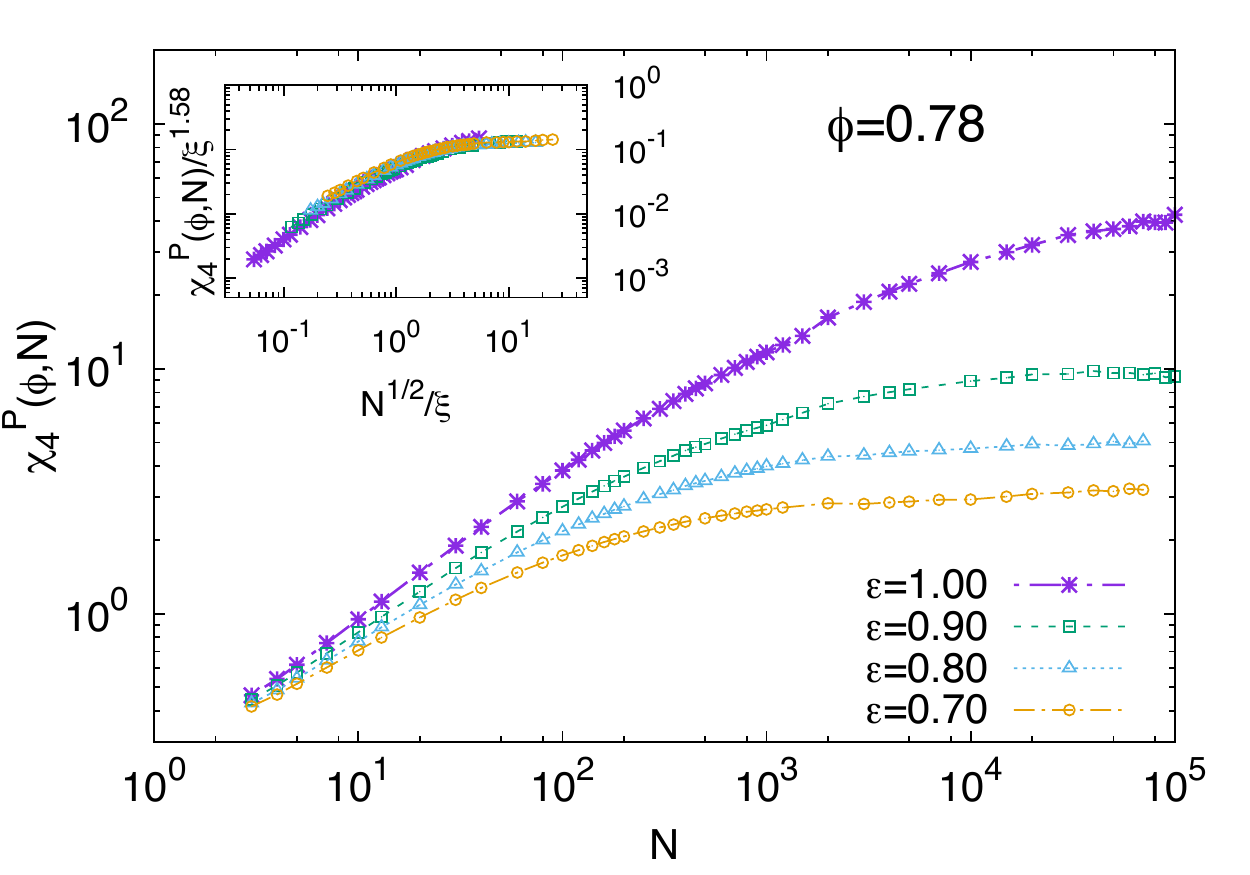}
  \end{center}
  \caption{Peak value of the dynamical susceptibility, $\chi_4^P$,
    versus $N$ for the packing fraction $\phi=0.78$ and coefficients
    of restitution $\varepsilon=1.0$, $0.90$, $0.80$,
    $0.70$. $\chi_4^P$ grows with $N$ before it saturates and it also
    grows as $\varepsilon$ is increased. Inset: the same data scaled
    as in Fig.~\ref{chi4P}(b).}
  \label{chi_4-peak}
\end{figure}
In Fig.~\ref{chi_4-peak} we present data for the peak value of
$\chi_4^P$ as a function of $N$. For a given $\phi$, here $\phi=0.78$,
the more inelastic systems are further away from criticality and hence
$\chi_4(t)$ and $\xi$ are smaller. However the data can be collapsed
with an $\varepsilon$-independent value of $d_f'=1.58$.

 All the scaling results described in this section show no 
significant difference between the elastic case $\varepsilon=1$ and 
the dissipative case $\varepsilon < 1$. Therefore,
we conclude that structural arrest occurs at higher packing fractions
for the more inelastic systems but the main characteristics of
dynamical heterogeneities are qualitatively the same for all investigated values of
$\varepsilon$.

\section{Conclusions}\label{conclusions}

 Via event driven simulations, we have investigated a
  two-dimensional homogeneously driven dissipative binary hard sphere
  system. All our results are consistent with a non-equilibrium glass
  transition which is controlled by the packing fraction,
  $\phi\to\phi_c$. Approaching this transition, we find long-lived
  and long-ranged dynamic heterogeneities which we have analyzed with
  several tools. Slow particles have been identified and analyzed in
terms of a statistical distribution of cluster sizes. The latter obeys
scaling and approaches an algebraic decay as $\phi\to\phi_c$,
suggesting that macroscopically large clusters of slow particles exist
in that limit.  Similarly, clusters of fast particles can be
identified. However, while the cluster size probability distributions
for slow clusters have a behavior that is reminiscent of percolating
systems near their critical point, the probability distributions for
the sizes of fast clusters have a non-critical log-normal
behavior. This suggests that any attempt at probing possible critical
phenomena associated with the dynamical arrest should focus on the
slow particles and not on the fast ones.

The spatial extent of the clusters has been characterized by the
radius of gyration. Relating cluster size and spatial extent reveals a
fractal dimension of the slow clusters which grows with $\phi$. In
other words, the slow particles aggregate into progressively more
compact clusters as $\phi\to\phi_c$.

Another route to studying dynamical heterogeneities is based on the overlap,
defined as the fraction of particles which have moved less than a
distance $a$ (usually $0.6$) in a given time interval t. The overlap
itself shows pronounced slowing down as evidenced by a strong increase
of the $\alpha$- relaxation time as $\phi\to\phi_c$. However, the
time-dependent overlap does not obey time-density superposition, 
similar to results in Ref.~\cite{FlennerSzamel2015} for a
2D non--dissipative fluid. In contrast, Abate and Durian's~\cite{Abate2007} 
data for the overlap Q(t) for a 2D air-fluidized
granular system seems to at least approximately satisfy time-density
superposition (Fig.~3, top panel in~\cite{Abate2007}). All our data
for different packing fractions and different coefficients of
restitution can nevertheless be collapsed to a single curve with help
of an empirical fit. The long-time decay is predicted to be algebraic
in time.

Of particular interest are fluctuations of the overlap, -- either
global ones as measured by $\chi_4(t)$ or spatially resolved ones
encoded in $S_4(q,t)$. The latter have been shown to obey scaling and
are well approximated by the Ornstein--Zernike form. This allows us to
extract a correlation length and the strength of the global
fluctuations $\lim_{q\to 0}S_4(q,t)$. Since the latter have been
computed independently, we can thereby show that $\chi_4(t)=\lim_{q\to
  0}S_4(q,t)$. This result relies on our sub-box analysis mimicking a
grand canonical ensemble with respect to particle number and
concentration. The four point susceptibility was previously shown to
diverge as $\chi_4(t)\propto (\phi_c-\phi)^{-\gamma_{\chi}}$, a result
that we associate with a diverging number of correlated
particles. Finite size scaling of $\chi_4(\phi,N)$ allows us
furthermore to relate cluster size and correlation length. Using the
previously determined values for $\xi$, we can collapse the data
approximately to a single curve, providing a consistency check for the
previously determined exponents $d_f',\gamma_{\xi}$ and
$\gamma_{\chi}.$

We have investigated the robustness of our results with respect to
variations in the cutoff $a$ and the coefficient of restitution
$\varepsilon$.  The results suggest that the geometry of the clusters
is largely insensitive to the definition of slow and fast particles
and to the degree of inelasticity of the collisions between the
particles. Even though $\chi_4$ and $\xi$ individually depend on $a$
and $\varepsilon$, % 
the relation $\chi_4(\tau_{\alpha})\propto \xi^{d_f'}(\tau_{\alpha})$
is surprisingly independent of those parameters.\\

\subsubsection*{Acknowledgments}
% \acknowledgments
We thank A. Fiege, I. Gholami
and T. Kranz for help with the numerical simulations. 
H.E.C. thanks E. Flenner, and G. Szamel for discussions. This work was
supported in part by DFG under grants SFB 602 and FOR 1394, by DOE
under grant DE-FG02-06ER46300, by NSF under grants PHY99-07949 and
PHY05-51164, and by Ohio University. 
K.E.A. acknowledges the CMSS program at Ohio University for partial support. 
K.V.L. thanks the Institute of 
Theoretical Physics, University of G\"ottingen, for hospitality and financial support.

\section*{Appendix}\label{appendix}
\begin{figure}[ht]
  {\includegraphics[scale=.65]{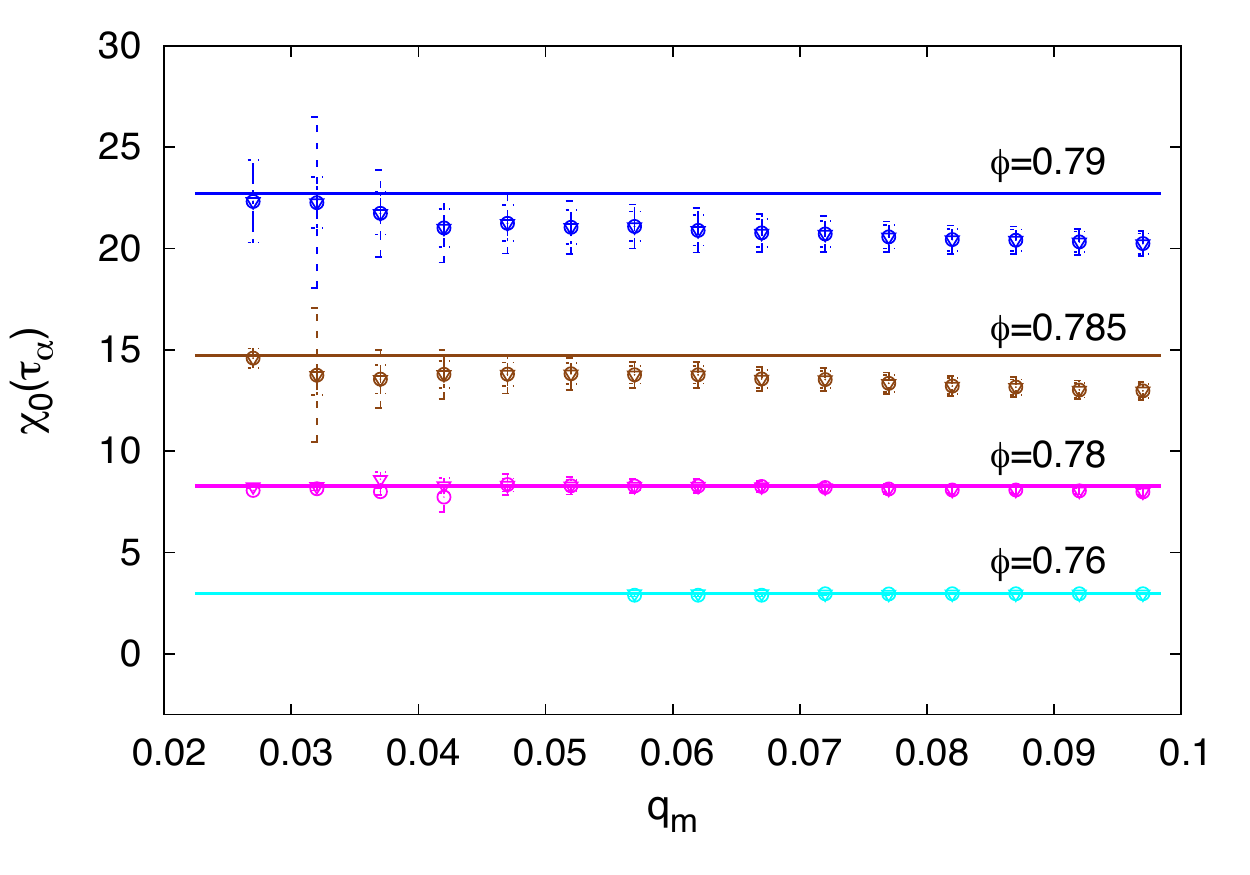}} 
   \caption{Comparison of
    $\chi_4(\tau_{\alpha})$ obtained from Eq.~(\ref{chi_4-eq}), shown by
    the solid line, to $\chi_0$ obtained by fitting functions
    (2) and (3), shown as triangles and circles, respectively. In almost all cases 
    the two values fall almost exactly on top of each other. The values of $\chi_0$ obtained
    from both fitting functions are very close to $\chi_4$.} 
    \label{fitting}
\end{figure} 
To check if the value of $\chi_4(\tau_{\alpha})$ can be obtained by an extrapolation of $S_4$ we determined
$\chi_0(\tau_{\alpha})=\lim_{q \rightarrow 0}S_4(q,\tau_{\alpha})$ by allowing for more
general fitting functions than just the OZ form:
\begin{equation}
S_4(q,\tau_{\alpha})=\frac{\chi_0}{1+({q}\xi)^2+A^2({q}\xi)^4}
\label{S4-fits}
\end{equation} 
suggested in~\cite{Flenner2011,Karmakar2010}. To confirm
Eq.~(\ref{qto0limit}) we have determined $\chi_0(t)$ from two different
fits of the above generalized expression for $S_4(q)$: in fit 2, $A=0$
and $\chi_0$ is a fit parameter and in fit 3, both $A$
and $\chi_0$ are fitted. The values obtained are shown as symbols in Fig.~\ref{fitting}, and are compared to
the value of $\chi_4(\tau_{\alpha})$ obtained form
Eq.~(\ref{chi_4-eq}), shown as the solid line in the same
figure. Furthermore we show in Fig.~\ref{fitting} the robustness 
of the fit with respect to the fitting range $[0:q_m]$. 
The agreement between $\chi_0(t)$ and $\chi_4(t)$ is
remarkably good -- for all packing fractions.

In addition, these different fits provide a check for the robustness
of the extracted values of the correlation length. We found that
different fitting functions did not change our results
significantly. This holds for the whole range of packing fractions. 

\renewcommand\refname{Notes and references}


\begin{thebibliography}{10}%

\bibitem{Ediger2000}
M.~D.~Ediger,
Annu. Rev. Phys. Chem.  
\textbf{51} 99-128 (2000).

 \bibitem{Berthier2011}
L.~Berthier, 
G.~Biroli, 
J.-P. Bouchaud, 
L.~Cipelletti, and
W. van Saarloos,
{\em Dynamical heterogeneities in glasses, colloids and granular materials} 
(Oxford University Press, Oxford, 2011).

\bibitem{Russell2000}
E.~V.~Russell, and
N.~E.~Israeloff,
Nature
 \textbf{408}, 695-698 (2000).

\bibitem{Dauchot2005}
O.~Dauchot,
G.~Marty, and
G.~Biroli,
Phys. Rev. Lett.
\textbf{95}, 265701 (2005).

\bibitem{Berthier2011-2}
L.~Berthier,
Physics 
\textbf{4}, 42 (2011).

\bibitem{Debenedetti2001}
P.~G.~Debenedetti, and 
F.~H.~Stillinger,
Nature 
\textbf{410}, 259-267 (2001).

\bibitem{Majmudar2007}
T.~S.~Majmudar,
M.~Sperl, 
S.~Luding, and 
R.~P.~Behringer,
Phys. Rev. Lett.
\textbf{98}, 058001 (2007).

\bibitem{Liu1998}
A.~Liu and S.~Nagel, 
Nature
\textbf{396}, 21 (1998).


\bibitem{Voigtmann2011}
T.~Voigtmann, 
Eur. Phys. J. E
\textbf{34}, 106 (2011).

\bibitem{Ikeda2012}
A.~Ikeda,
L.~Berthier and 
P.~Sollich, 
Phys. Rev. Lett.
\textbf{109}, 018301 (2012).

\bibitem{Abate2007}
A.~R.~Abate and 
D.~J.~Durian, 
Phys. Rev. E
\textbf{76}, 021306 (2007).

\bibitem{AvilaPRL2014}
K.~E.~Avila,
H.~E.~Castillo,
A.~Fiege,
K.~Vollmayr-Lee, and
A.~Zippelius,
Phys. Rev. Lett.
\textbf{113}, 025701 (2014).

\bibitem{Gholami2011}
I.~Gholami,
A.~Fiege, and
A.~Zippelius,	
Phys. Rev. E 
\textbf{84}, 031305 (2011).


\bibitem{Lechenault2008} 
F.~Lechenault, O.~Dauchot, G.~Biroli, and J.~P.~Bouchaud,
Europhys. Lett. \textbf{83}, 46003 (2008). 

\bibitem{Wortel2014}
G.~H.~Wortel, J.~A.~Dijksman, and M.~van~Hecke, Phys. Rev. E
\textbf{89}, 012202 (2014). 

\bibitem{Keys2007}%
A.~S. Keys,
A.~R. Abate,
S.~C. Glotzer, and
D.~J. Durian,
Nature Physics
\textbf{3}, 260-264 (2007).

\bibitem{Espanol1995}
  P. Espanol, and P. Warren, Europhys. Lett.  \textbf{30} 191 (1995).

\bibitem{Fiege2009} 
  I.~Gholami, T.~Aspelmeier, and A. Zippelius,  
  Phys. Rev. Lett.,
  \textbf{102}, 098001 (2009). 
  

\bibitem{Bengtzelius1984} 
  U. Bengtzelius, W. G\"otze, and A. Sjolander,  
  J. Phys. C,
  \textbf{17}, 5915 (1984). 
  
  \bibitem{Kranz2010}
W. T. Kranz, 
M. Sperl, and 
A. Zippelius, 
Phys. Rev. Lett. 
\textbf{104}, 225701 (2010); 
Phys. Rev. E 
\textbf{87}, 022207 (2013).


  \bibitem{Brambilla2009}
G.~Brambilla, 
D.~El~Masri, 
M.~Pierno, 
L.~Berthier,
L.~Cipelletti, 
G.~Petekidis, and 
A.~B.~Schofield, 
Phys. Rev. Lett. 
\textbf{102}, 085703 (2009).

\bibitem{Berthier2009}
L. Berthier, and T. A. Witten,
 Europhys. Lett., \textbf{86}, 10001 (2009).

\bibitem{FlennerSzamel2015}
E.~Flenner, and
G.~Szamel,
Nat. Commun. 
\textbf{6}, 7392 (2015).


\bibitem{Weeks2000}
E.~R.~Weeks, 
J.~C.~Crocker, 
A.~C.~Levitt, 
A.~Schofield, and 
D.~A.~Weitz,
Science 
\textbf{287} 627-631 (2000).

\bibitem{Parsaeian2009}
A.~Parsaeian, and
H.~E.~Castillo,
Phys. Rev. Lett. 
\textbf{102}, 055704 (2009).

\bibitem{Chaudhuri}
P.~Chaudhuri,
L.~Berthier, and
W.~Kob,
Phys. Rev. Lett.
\textbf{99}, 060604 (2007).

\bibitem{Sengupta2013}
S.~Sengupta, 
S.~Karmakar, 
C.~Dasgupta, and 
S.~Sastry,
J. Chem. Phys. 
{\bf 138}, 12A548 (2013).

 \bibitem{footnote1}
 In the case of the fast clusters, we do not 
have sufficient statistics to determine $d_f$.

\bibitem{Villarica1993}
M. Villarica, M. J. Casey, J. Goodisman and J. Chaiken
J. Chem. Phys. {\bf 98}, 4610 (1993).

\bibitem{Wang1994}
Chun-Ru Wang, Rang-Bin Huang, Zhao-Yang Liu, Lan-Sun Zheng, Chemical
Physics Letters {\bf 227}, 103 (1994).

\bibitem{Mendham2001}
J. Mendham, N. Hay, M. B. Mason, J. W. G. Tisch, and J. P. Marangos, 
Phys. Rev. A {\bf 64}, 055201 (2001).

\bibitem{Buhrman1976}
R. A. Buhrman and C. G. Granqvist, J. Appl. Phys., {\bf 47}, 2220,
(1976). 

\bibitem{Vesperini2000}
E. Vesperini, Mon. Not. R. Astron. Soc. {\bf 318}, 841, (2000).

\bibitem{Toninelli2005}
C.~Toninelli,
M.~Wyart,
L.~Berthier,
G.~Biroli, and
J-P.~ Bouchaud,
Phys. Rev. E
\textbf{71}, 041505 (2005).

%\bibitem{Doliwa2000}
%B. Doliwa, and A. Heuer, 
%Phys. Rev. E \textbf{61}, 6898 (2000).

\bibitem{Parsaeian2008}
A. Parsaeian, and H.~E. Castillo
Phys. Rev. E \textbf{78}, 060105 (R) (2008).

\bibitem{Flenner2011}
E.~Flenner,
M.~Zhang, and
G.~Szamel,
Phys. Rev. E 
\textbf{83}, 051501 (2011).

\bibitem{footnote2}
In~\cite{AvilaPRL2014} we used a
  slightly different definition of $Q_{\bf r}(t;t_0)$: we normalized
  by $N_{\bf r}$ instead of $N$. Consequently the fluctuations
  vanished for small times, when averaged over all space. With the
  present definition the fluctuations remain finite even at $t=0$. The
  important fluctuations at large times are unaffected by the
  normalization.

\bibitem{Lebowitz}
J.~L.~Lebowitz, 
J.~K.~Percus, and 
L.~Verlet, 
Phys. Rev. 
\textbf{153}, 250 (1967).

\bibitem{Berthier2005}
 L.~Berthier, 
G.~Biroli, 
J.-P. Bouchaud, 
L.~Cipelletti,
D.~Masri,
D.~L'Hôte,
F.~Ladieu, and
M.~Pierno,
Science 
\textbf{310}, 1797 (2005).

\bibitem{Glotzer2003}
N.~La\v{c}evi\'c, 
F.~W.~Starr,
T.~B. Schr\o der, and
S.~C.~Glotzer, 
J. Chem. Phys.
\textbf{119}, 7372 (2003).

\bibitem{footnote3} 
Notice that the exponents $\gamma_{\chi}$ and $\gamma_{\xi}$ have been
determined from results at $t = \tau_{\alpha}$, but $\chi^P_4 =
\chi_4(\tau_4)$, with $\tau_4$ generally a longer time than
$\tau_{\alpha}$. Despite this, there is good data collapse in
Fig.~\ref{chi4P}(b) and in the inset of Fig.~\ref{chi_4-peak}. 

\bibitem{Karmakar2009}
S.~Karmakar,
C.~Dasgupta, and
S.~Sastry,
PNAS
\textbf{106}, 3677 (2009).
 
     \bibitem{Karmakar2010} 
S.~Karmakar, C.~Dasgupta, and S.~Sastry,
  Phys. Rev. Lett.  \textbf{105}, 015701 (2010).

\end{thebibliography}
\end{document}